\documentclass[]{JHEP3}

\usepackage{amsmath}
\usepackage{amsfonts}
\usepackage{amssymb}
\usepackage{epsfig}

\usepackage{epsfig,multicol,bbm}

\title{Energy-independent new physics in the flavour ratios of high-energy astrophysical neutrinos}

\author{
M.~Bustamante$^{a,b}$,~A.M.~Gago$^a$,~C.~Pe\~na-Garay$^c$ \\
$^a$Secci\'on F\'isica, Departamento de Ciencias, Pontificia
Universidad Cat\'olica del Per\'u, Apartado 1761, Lima,
Peru \\
$^b$Theoretical Physics Department, Fermi National Accelerator Laboratory, Batavia, IL 60510, USA \\
$^c$Instituto de F\'isica Corpuscular (IFIC), Centro Mixto CSIC-UVEG Edificio Investigaci\'on Paterna, Apartado 22085, 46071 Valencia, Spain \\
E-mail: \email{mbustamante@pucp.edu.pe,agago@pucp.edu.pe,carlos.penya@ific.uv.es}
}

\preprint{\hepph{1001.4878}}

\abstract{We have studied the consequences of breaking the CPT symmetry in the neutrino sector, using the expected high-energy neutrino flux from distant cosmological sources such as active galaxies. For this purpose we have assumed three different hypotheses for the neutrino production model, characterised by the flavour fluxes at production $\phi_e^0:\phi_\mu^0:\phi_\tau^0 = 1:2:0$, $0:1:0$, and $1:0:0$, and studied the theoretical and experimental expectations for the muon-neutrino flux at Earth, $\phi_\mu$, and for the flavour ratios at Earth, $R = \phi_\mu/\phi_e$ and $S = \phi_\tau/\phi_\mu$. CPT violation (CPTV) has been implemented by adding an energy-independent term to the standard neutrino oscillation Hamiltonian. This introduces three new mixing angles, two new eigenvalues and three new phases, all of which have currently unknown values. We have varied the new mixing angles and eigenvalues within certain bounds, together with the parameters associated to pure standard oscillations. Our results indicate that, for the models $1:2:0$ and $0:1:0$, it might possible to find large deviations for $\phi_\mu$, $R$, and $S$ between the cases without and with CPTV, provided the CPTV eigenvalues lie within $10^{-29}\--10^{-27}$ GeV, or above. Moreover, if CPTV exists, there are certain values of $R$ and $S$ that can be accounted for by up to three production models. If no CPTV were observed, we could set limits on the CPTV eigenvalues of the same order. Detection prospects calculated using IceCube suggest that for the models $1:2:0$ and $0:1:0$, the modifications due to CPTV are larger and more clearly separable from the standard-oscillations predictions. We conclude that IceCube is potentially able to detect CPTV but that, depending on the values of the CPTV parameters, there could be a mis-determination of the neutrino production model.}


\begin{document}

\section{Introduction}\label{Section_Intro}

Experiments performed over the last thirty years have established that neutrinos can change flavour: careful measurements of solar \cite{Davis:1968cp,Aharmim:2005gt,Arpesella:2008mt}, atmospheric \cite{Ashie:2004mr,Ambrosio:2004ig}, reactor \cite{Eguchi:2002dm} and accelerator \cite{Ahn:2006zza,Adamson:2008zt} neutrinos have established that there is a nonzero probability that a neutrino created with a certain flavour is detected with a different one after having propagated for some distance, and that this probability is a periodic function of the propagated distance, $L$, and the neutrino energy, $E$. The standard mechanism of neutrino flavour change requires neutrinos to be massive and results in a probability of flavour change that is oscillatory, with oscillation lengths that have a distinct $1/E$ dependence (see Section \ref{Section_Theory_Sub_StdOsc}).
 
So far, the experiments that have studied neutrino flavour transitions \cite{Walter:2008ar} have been designed to detect neutrinos with energies that range from a few MeV (solar neutrinos) to the TeV scale (atmospheric neutrinos). Notably, data from the Super-Kamiokande atmospheric neutrino experiment \cite{Fogli:1999fs} was used to find an energy dependence of the oscillation probability of $E^n$, with $n = -0.9 \pm 0.4$ at $90\%$ confidence level, thus confirming the dominance of the mass-driven mechanism in this energy range, and relegating any other potential mechanisms to subdominance. It is possible, however, that one or more of such subdominant mechanisms become important at higher energies. 

In the present paper, we have explored a possible scenario where there is an additional oscillation mechanism present which results in an energy-independent contribution to neutrino oscillations. This mechanism, though subdominant in the MeV\---TeV range, might become dominant at higher energies, where the $1/E$ dependence of the standard oscillation term might render it comparatively unimportant: the higher the energy, the stronger the suppression of the standard oscillation term. The highest-energy flux of neutrinos available is the expected ultra-high-energy (UHE, with energies at the PeV scale and higher) flux from astrophysical sources \---notably, active galaxies (see Section \ref{Section_IceCube})\--- which are located at distances in the order of tens or hundreds of Mpc. 

We will deferr the detailed treatment of how the energy-independent contribution is introduced to Section \ref{Section_Theory} and focus now on the possible mechanisms that might motivate it. According to the CPT theorem, any Lorentz-invariant local quantum field theory must be built out of a CPT-conserving Lagrangian. However, the Standard Model (SM) is known to be valid at energies well below the Planck scale, $m_\text{Pl} \simeq 10^{19}$ GeV and, at higher energies, motivated by theories beyond the SM \cite{Mavromatos:2004ug,Mavromatos:2009ww}, CPT and Lorentz invariance might be broken. At accessible energies, the breaking of these symmetries can be described by an effective field theory that contains the SM. We have explored the possibility that CPT is not an exact symmetry, but rather that it is broken by the addition of a CPT-odd term to an otherwise CPT-even Lagrangian. The observation of the non-conservation of CPT would imply a fundamental revision of the usefulness of local quantum field theories as accurate descriptions of fundamental interactions. A possible realisation of a CPT-violating (CPTV) effective field theory is the Standard Model Extension \cite{Colladay:1998fq,Kostelecky:2003xn}, which contains the SM, conserves $SU\left(3\right) \times SU\left(2\right) \times U\left(1\right)$, and also considers potential Lorentz- and CPT-violating couplings in the gauge, lepton, quark and Yukawa sectors. It is worth noting that an alternative mechanism that also results in an energy-independent contribution to the oscillations is the nonuniversal coupling of the different neutrino flavours to an external torsion field \cite{DeSabbata:1981ek}.

The rest of the paper is organised as follows. In Section \ref{Section_Theory}, we explain how the contributions from potential energy-independent new physics affect the flavour-transition probability and introduce the detected fluxes of the different flavours of astrophysical neutrinos, and the ratios between them, as our observables. Section \ref{Section_Effect} discusses the deviations of the flavour ratios from their standard values in three different cases: when only $\nu_\mu$ can be detected, when $\nu_\mu$ and $\nu_e$ can be detected, and when the three flavours, $\nu_e$, $\nu_\mu$, $\nu_\tau$, can be detected\footnote{Some preliminary results on the three-flavour detection case were presented by the authors elsewhere \cite{Bustamante:2009wz}.}. In Section \ref{Section_IceCube}, we predict the feasibility of detecting a possible energy-independent contribution using the IceCube neutrino detector, and larger-volume versions of it. We summarise our results and conclude in Section \ref{Section_Conclusions}.

\section{Theoretical framework}\label{Section_Theory}

\subsection{Standard mass-driven flavour oscillations}\label{Section_Theory_Sub_StdOsc}

The standard mechanism that explains neutrino flavour transitions makes use of two different bases: the basis of neutrino mass eigenstates, which have well-defined masses, and the basis of neutrino interaction states \---the flavour basis\--- which are the ones that take part in weak processes such as $W$ decay. The two bases are connected through a unitary transformation, so that we can write each one of the flavour states $\lvert\nu_\alpha\rangle$ as a linear combination of the mass eigenstates $\lvert\nu_i\rangle$, i.e.,
\begin{equation}\label{EqFlavStateDef}
 \lvert \nu_\alpha \rangle = \sum_i \left[U_0\right]_{\alpha i}^\ast \lvert \nu_i \rangle \ ,
\end{equation}
where the coefficients $\left[U_0\right]_{\alpha i}$ are components of the unitary mixing matrix that represents the transformation. Assuming the existence of three active neutrino families ($\alpha = e,\mu,\tau$), as indicated by LEP \cite{Decamp:1989tu}, three mass eigenstates are required ($i=1,2,3$ in the sum) and so $U_0$ is a $3\times3$ matrix. The mass eigenstates $\lvert \nu_i \rangle$ satisfy Schr\"{o}dinger's equation and so propagate, sans a flavour-independent common phase $e^{-iEL}$, as
\begin{equation}
 \lvert \nu_i\left(L\right) \rangle = e^{-iHL} \lvert \nu_i \rangle = e^{-i\frac{m_i^2}{2E}L} \lvert \nu_i \rangle \ ,
\end{equation}
where we have assumed that $m_i \ll E$, so that $p = \sqrt{E^2-m_i^2} \simeq E - m_i^2/(2E)$, and that, because neutrinos are highly relativistic particles, $t \simeq L$ (in natural units). Thus, a neutrino created with a definite flavour $\alpha$ will, in general, become a superposition of states of different flavour as it propagates, so that a detector placed in its way can, with a certain probability, register it as having a different flavour from the one that it was created with. This becomes evident by writing the Hamiltonian in the basis of flavour eigenstates, a choice which will also allow us to introduce contributions from new physics later in a more straightforward manner. In this basis, if a neutrino is produced with flavour $\alpha$, then, after having propagated for a distance $L$, its evolved state will be 
\begin{equation}\label{EqNuEvolved}
 \lvert \nu_\alpha\left(L\right) \rangle = e^{-iH_mL} \lvert \nu_\alpha \rangle \ ,
\end{equation}
where the oscillation Hamiltonian $H_m$ is the one corresponding to the standard, mass-driven, mechanism, and is written in the flavour basis. $H_m$ is related to the Hamiltonian in the mass basis \--the ``mass matrix''\-- through a similarity transformation that makes use of the unitary mixing matrix $U_0$:
\begin{equation}\label{EqHm}
 H_m = U_0 H U_0^\dag = U_0 \frac{\text{diag}\left(0,\Delta m_{21}^2,\Delta m_{31}^2\right)}{2E} U_0^\dag \ .
\end{equation}
$U_0$ is the Pontecorvo-Maki-Nakagawa-Sakata (PMNS) mixing matrix, which in the PDG parametrisation \cite{Amsler:2008zzb} can be written in terms of three mixing angles, $\theta_{12}$, $\theta_{13}$ and $\theta_{23}$, and one CP-violation phase, $\delta_\text{CP}$, as
\begin{eqnarray*}\label{EqPMNSMatrix}
 U_0\left(\left\{\theta_{ij}\right\}, \delta_\text{CP}\right)
 = \left(\begin{array}{ccc}
     c_{12}c_{13} & s_{12}c_{13} & s_{13}e^{-i\delta_\text{CP}} \\
     -s_{12}c_{23}-c_{12}s_{23}s_{13}e^{i\delta_\text{CP}} & c_{12}c_{23}-s_{12}s_{23}s_{13}e^{i\delta_\text{CP}} & s_{23}c_{13} \\
     s_{12}s_{23}-c_{12}c_{23}s_{13}e^{i\delta_\text{CP}} & -c_{12}s_{23}-s_{12}c_{23}s_{13}e^{i\delta_\text{CP}} & c_{23}c_{13}
 \end{array}\right) ,
\end{eqnarray*}
with $c_{ij} \equiv \cos\left(\theta_{ij}\right)$, $s_{ij} \equiv \sin\left(\theta_{ij}\right)$. The standard flavour-oscillation probability $P_{\alpha\beta} = \left\vert \langle \nu_\beta \vert \nu_\alpha\left(L\right) \rangle \right\vert^2$ can hence be calculated, using Eq.~(\ref{EqNuEvolved}) for the evolved neutrino state and Eq.~(\ref{EqHm}) for the standard Hamiltonian, to be
\begin{equation}\label{EqTransProb}
 P_{\alpha\beta} 
 = \delta_{\alpha\beta} - 4\sum_{i>j} \text{Re}\left(J_{\alpha\beta}^{ij}\right) \sin^2\left(\frac{\Delta m_{ij}^2}{4E}L\right)
 + 2 \sum_{i>j} \text{Im}\left(J_{\alpha\beta}^{ij}\right) \sin\left(\frac{\Delta m_{ij}^2}{2E}L\right) ,
\end{equation}
where $\Delta m_{ij}^2 \equiv m_i^2 - m_j^2$, with $m_i$ the mass of the $i$-th eigenstate, and 
\begin{equation}
 J_{\alpha\beta}^{ij} \equiv \left[U_0\right]_{\alpha i}^\ast \left[U_0\right]_{\beta i} \left[U_0\right]_{\alpha j} \left[U_0\right]_{\beta j}^\ast \ .
\end{equation}
(For a detailed deduction, see, e.g., \cite{Kayser:2008rd}.) It is straightforward to conclude from Eq.~(\ref{EqTransProb}) that flavour transitions occur because neutrinos are massive, particularly, because different mass eigenstates have different masses (clearly, if $\Delta m_{ij}^2 = 0$, no transitions occur), and because flavour states are not mass eigenstates. Note the $1/E$ dependence on the energy associated with this standard, mass-driven, oscillation mechanism. Note also that the full form of the PMNS matrix includes two extra Majorana CP-violation phases, $\alpha_1$ and $\alpha_2$, which are identically zero if neutrinos are Dirac, so that the complete matrix is given by
\begin{equation}
 U_0 \times \text{diag}\left(e^{i\alpha_1/2},e^{i\alpha_2/2},1\right) .
\end{equation}
However, these phases do not affect the oscillations, so we have not included them in the definition of the standard mixing matrix $U_0$. 

Because, as was mentioned in Section \ref{Section_Intro}, we will be considering UHE neutrinos of extragalactic origin, the flavour-transition probability in Eq.~(\ref{EqTransProb}) oscillates very rapidly, and we use instead the average probability, which is obtained by averaging the oscillatory terms in the expression, thus yielding
\begin{equation}\label{EqTransProbAvg}
 \langle P_{\alpha\beta} \rangle = \sum_i \lvert \left[U_0\right]_{\alpha i}\rvert^2 \lvert \left[U_0\right]_{\beta i}\rvert^2 \ .
\end{equation}
When using the average probability, the information in the oscillation phase, including any potential CPTV energy-independent contribution, is lost. The modifications to the mixing angles due to CPTV, however, will still be present in the averaged version of the probability.

Using the latest data from solar, atmospheric, reactor (KamLAND and CHOOZ) and accelerator (K2K and MINOS) experiments, the authors of \cite{Schwetz:2008er} performed a global three-generations fit and found the best-fit values of the standard oscillation parameters and their $3\sigma$ intervals to be 
\begin{equation}\label{EqStdMixPar}
 \Delta m_{21}^2 = 7.65^{+0.69}_{-0.60} \times 10^{-5} ~\text{eV}^2 \ , \qquad
 \lvert \Delta m_{31}^2 \rvert = 2.40^{+0.35}_{-0.33} \times 10^{-3} ~\text{eV}^2  \,
\end{equation}
\begin{equation}
 \sin^2\left(\theta_{12}\right) = 0.304^{+0.066}_{-0.054} \ , \qquad
 \sin^2\left(\theta_{13}\right) = 0.01^{+0.046}_{-0.01} \ , \qquad
 \sin^2\left(\theta_{23}\right) = 0.50^{+0.17}_{-0.14} \ . 
\end{equation}
The are no experimental values for $\delta_\text{CP}$ presently. We have assumed a normal mass hierarchy, so that $\Delta m_{32}^2 = \Delta m_{31}^2 - \Delta m_{21}^2$.

\subsection{Adding an energy-independent Hamiltonian}\label{Section_Theory_Sub_EnergyInd}

The lepton sector of the Standard Model Extension contains the Lorentz-violating contributions \cite{Colladay:1998fq}
\begin{equation}
 \mathcal{L}_\text{LIV} \supset - \left(a_L\right)_{\mu\alpha\beta} \overline{L}_\alpha \gamma^\mu L_\beta
                     + \frac{1}{2} i \left(c_L\right)_{\mu\nu\alpha\beta} \overline{L}_\alpha \gamma^\mu \overleftrightarrow{D}^\nu L_\beta \ ,
\end{equation}
where $L_\alpha$ is the usual SM lepton doublet and $\alpha$, $\beta$ are flavour indices. The first term is CPT-odd, with the coefficients $a_L$ having dimensions of mass, while the second term is CPT-even, with the $c_L$ dimensionless. Because we are interested in an energy-independent CPTV contribution, we have kept only the first term in the Lagrangian. In the neutrino sector, then, CPT violation can be introduced through an effective, model-independent, vector coupling of the form \cite{Dighe:2008bu}
\begin{equation}\label{EqLagrangian}
 \mathcal{L}_\text{CPTV}^\nu = \overline{\nu}^\alpha b_\mu^{\alpha\beta} \gamma^\mu \nu^\beta \ .
\end{equation}
The vector $\overline{\nu}^\alpha \gamma^\mu \nu^\beta$ is CPT-odd and $b_\mu^{\alpha\beta}$ are real coefficients, so $\mathcal{L}_\text{CPTV}^\nu$ is CPT-odd, i.e., $\text{CPT}\left(\mathcal{\mathcal{L}_\text{CPTV}^\nu}\right) = - \mathcal{\mathcal{L}_\text{CPTV}^\nu}$. When the effective energy-independent Hamiltonian associated to $\mathcal{L}_\text{CPTV}^\nu$ is added to the standard mass-driven neutrino oscillation Hamiltonian, it modifies the energy eigenvalues and, as a result, the mixing matrix is modified as well.

Motivated by the vector coupling in Eq.~(\ref{EqLagrangian}), and in analogy to the standard oscillation scenario, we can introduce an energy-independent contribution in the form of the Hamiltonian (also in the flavour basis)
\begin{equation}
 H_b = U_b ~\text{diag}\left(0,b_{21},b_{31}\right) U_b^\dag \ ,
\end{equation}
where $b_{ij} \equiv b_i - b_j$. Following \cite{Dighe:2008bu}, we write the mixing matrix in this case as
\begin{equation}
 U_b = \text{diag}\left(1,e^{i\phi_2},e^{i\phi_3}\right)U_0\left(\left\{\theta_{bij}\right\}, \delta_b\right) .
\end{equation}
The mixing angles associated with this Hamiltonian are $\theta_{b12}$, $\theta_{b13}$, $\theta_{b23}$, and $\delta_b$ fills the role of $\delta_\text{CP}$ in the standard Hamiltonian. The two extra phases, $\phi_2$ and $\phi_3$, appear because, once the flavour states and the mass eigenstates have been related through Eq.~(\ref{EqFlavStateDef}), the former are completely defined, and the two extra phases cannot be rotated away.

$H_b$ is dependent on eight parameters \---two eigenvalues ($b_{21}$, $b_{31}$), three mixing angles ($\theta_{b12}$, $\theta_{b13}$, $\theta_{b23}$) and three phases ($\delta_b$, $\phi_2$, $\phi_3$)\--- whose values are currently unknown. There are, however, experimental upper limits \cite{Dighe:2008bu} on $b_{21}$, obtained using solar and Super-Kamiokande data, and on $b_{32}$, obtained using atmospheric and K2K data:
\begin{equation}\label{EqbijLimits}
 b_{21} \le 1.6 \times 10^{-21} ~\text{GeV} \ , \qquad b_{32} \le 5.0 \times 10^{-23} ~\text{GeV} \ .
\end{equation}

The full Hamiltonian, including standard oscillations and the energy-independent contribution, is then
\begin{equation}
 H_f = H_m + H_b \ .
\end{equation}
In Section \ref{Section_Intro}, we saw that $H_m$ has been experimentally demonstrated to be the dominant contribution to the oscillations in the low to medium energy (MeV\---TeV) regime: there are no indications of new energy-independent physics at these energies and accordingly the limits on $b_{ij}$ shown in Eq.~(\ref{EqbijLimits}) were placed. Because of the $1/E$ dependence of $H_m$, however, it remains possible that, at higher energies, where the contribution of $H_m$ is suppressed, the effect of a hypothetical energy-independent term $H_b$ becomes comparable to it or even dominant. Such energy requirement is expected to be fulfilled by the UHE astrophysical neutrino flux (see Section \ref{Section_Intro}).

We would like to write the flavour transition probability corresponding to this Hamiltonian in a form analogous to Eq. (\ref{EqTransProbAvg}). In order to do this, we need to know what is the mixing matrix $U_f$ that connects the flavour basis and the basis in which $H_f$ is diagonal. Using basic linear algebra, this is achieved simply by diagonalising $H_f$, finding its normalised eigenvectors, and building $U_f$ by arranging them in column form. The components of the resulting matrix are in general complicated functions of the standard mixing parameters ($\left\{\theta_{ij}\right\}$, $\left\{\Delta m_{ij}^2\right\}$, $\delta_\text{CP}$) and of the parameters of $H_b$ ($\left\{\theta_{bij}\right\}$, $\left\{b_{ij}\right\}$, $\delta_{b}$, $\phi_2$, $\phi_3$). By comparing the mixing matrix obtained by diagonalisation of $H_f$ with a general PMNS matrix, given explicitly by Eq.~(\ref{EqPMNSMatrix}) with mixing angles $\Theta_{ij}$ and phase $\delta_f$, we are then able to calculate how the effective mixing angles $\Theta_{ij}$ vary with the parameters of $H_b$ and $\delta_\text{CP}$. Succintly put, we have
\begin{equation}
 U_f 
 = U_f\left(\left\{\theta_{ij}\right\},\left\{\theta_{bij}\right\},\left\{\Delta  m_{ij}^2\right\},\left\{b_{ij}\right\},\delta_\text{CP},\delta_{b},\phi_{b2},\phi_{b3}\right) 
 = U_0\left(\left\{\Theta_{ij}\right\},\delta_f\right) \ .
\end{equation}
Note that we have not used a perturbative expansion in the $b_{ij}$, as in \cite{Dighe:2008bu}, to calculate $U_f$. This was done in order to allow for the possibility that the new physics effects become dominant at high energies, a possibility that would be negated if we had assumed that the effects are small from the start. As a result, the functional forms of the $\Theta_{ij}$ and $\delta_f$, while calculated in a straightforward manner, result in lengthy expressions and, due to their unilluminating character, we have chosen not to present them here.

Thus defined, $U_f$ is $14$-parameter function. However, the standard mixing parameters $\Delta m_{21}^2$, $\Delta m_{31}^2$, $\theta_{12}$, $\theta_{13}$ and $\theta_{23}$ have been fixed by neutrino oscillation experiments (see Eq.~(\ref{EqStdMixPar})). Additionally, in order to simplify the analysis, we have set the phases $\delta_b = \phi_2 = \phi_3 = 0$. The standard CP-violating phase $\delta_\text{CP}$ has been allowed to vary in the range $\left[0,2\pi\right]$ in some of our plots, but otherwise we have set it to zero as well. As a further simplification, we have made the eigenvalues of $H_m$ proportional to those of $H_b$, at a fixed energy of $E^\star = 1$ PeV, that is,
\begin{equation}
 b_{ij} = \lambda \frac{\Delta m_{ij}^2}{2E^\star} \ ,
\end{equation}
with $\lambda$ the proportionality constant. The upper bounds on the $b_{ij}$, Eq.~(\ref{EqbijLimits}), are satisfied for $\lambda \lesssim 10^4$. Standard, purely mass-driven oscillations are recovered when $\lambda = 0$. Thus we are left with only four free parameters to vary: $\lambda$, $\theta_{b12}$, $\theta_{b13}$ and $\theta_{b23}$ (and $\delta_\text{CP}$, where noted).

In analogy to Eq. (\ref{EqTransProbAvg}), the average flavour transition probability associated to the full Hamiltonian $H_f$ is then
\begin{equation}\label{EqTransProbAvgFull}
 \langle P_{\alpha\beta} \rangle = \sum_i \lvert\left[U_f\right]_{\alpha i}\rvert^2 \lvert\left[U_f\right]_{\beta i}\rvert^2 \ .
\end{equation}
We will use this expression for the flavour-transition probability hereafter. It is worth noting that, if the CPTV contribution were introduced instead through a modified energy-momentum relation, only the oscillation phase would be affected, and this information would be lost when the average probability was used in place of the oscillatory one \cite{Bazo:2009en}.

\subsection{Astrophysical neutrino flavour ratios}\label{Section_Theory_Sub_FlavRatios}

We have seen that, in order for a potential energy-independent contribution to the flavour transitions to be visible, we would need to use the expected UHE astrophysical neutrino flux. As mentioned in Section \ref{Section_Intro}, the sources of this flux, e.g., active galaxies, are located at distances of tens to hundreds of Mpc, so that the average flavour transition probability, Eq.~(\ref{EqTransProbAvgFull}), can be used.

If, at the sources, neutrinos of different flavours are produced in the ratios $\phi_e^0:\phi_\mu^0:\phi_\tau^0$, then, because of flavour transitions during propagation, the ratios at detection will be
\begin{equation}\label{EqFluxDetected}
 \phi_\alpha = \sum_{\beta=e,\mu,\tau} \langle P_{\beta\alpha} \rangle \phi_\beta^0 \ ,
\end{equation}
for $\alpha = e, \mu, \tau$. Note that the ratio in Eq.~(\ref{EqFluxDetected}) is the proportion of $\nu_\alpha$ to the sum of all flavours detected at Earth. We will later denote the actual neutrino fluxes, in units of GeV$^{-1}$ cm$^{-2}$ s$^{-1}$ sr$^{-1}$, by $\Phi_\alpha$. Evidently, the initial flavour ratios depend on the astrophysics at the source, which is currently not known with high certainty, while the detected ratios depend also on the oscillation mechanism and could be affected by the presence of an energy-independent contribution at high energies. Thus, the reconstruction of the initial neutrino fluxes from the detected ones is a difficult task \cite{Esmaili:2009dz,Barenboim:2003jm,Beacom:2003nh,Athar:2005wg,Lipari:2007su,Lai:2009ke,Xing:2006uk}. The effect of CPTV on the flavour ratios of high-energy astrophysical neutrinos has been explored elsewhere literature: in \cite{Barger:2000iv}, for instance, a two-neutrino approximation was employed and it was assumed that $H_m$ and $H_b$ are diagonalised by the same mixing matrix, i.e., that $\theta_{bij} = \theta_{ij}$, while, in \cite{Barenboim:2003jm}, neutrinos and antineutrinos were treated differently due to CPTV. Ref.~\cite{Dighe:2008bu} used a formalism similar to the one we have used, but applied it to long-baseline terrestrial experiments and, due to the lower energies involved, introduced the CPTV effects as perturbations. The main difference between the existing literature on the effects of CPTV on the flavour fluxes of high-energy astrophysical neutrinos and the present work is that we have not treated the CPTV contribution as a perturbation, but, rather, we have allowed for the possibility that it becomes dominant at a high enough energy scale.

The most commonly used assumption for the initial neutrino flux \cite{Athar:2005wg} considers that the charged pions created in high-energy proton-proton and proton-photon collisions decay into neutrinos and muons, which in turn decay into neutrinos: 
\begin{equation}
 \pi^+ \rightarrow \mu^+ \nu_\mu \rightarrow e^+ \nu_e \overline{\nu}_\mu \nu_\mu \ , \qquad
 \pi^- \rightarrow \mu^- \overline{\nu}_\mu \rightarrow e^- \overline{\nu}_e \nu_\mu \overline{\nu}_\mu \ .
\end{equation}
Such process yields approximately $\phi_e^0:\phi_\mu^0:\phi_\tau^0 = 1:2:0$ (see \cite{Lipari:2007su} for a more detailed treatment), where we have not discriminated between neutrinos and antineutrinos, as is the case with current \u{C}erenkov-based neutrino telescopes. In the standard oscillation scenario, i.e., in the absence of an energy-independent contribution, plugging this initial flux into Eq.~(\ref{EqFluxDetected}), and using the best-fit values of the mixing angles, Eq.~(\ref{EqStdMixPar}), results in equal detected fluxes of each flavour, i.e., $\phi_e^\text{std}:\phi_\mu^\text{std}:\phi_\tau^\text{std} \approx 1:1:1$. 

In a related production process \cite{Lipari:2007su,Rachen:1998fd,Kashti:2005qa}, the muons produced by pion decay lose most of their energy before decaying, so that a pure-$\nu_\mu$ flux is generated at the source, i.e., $\phi_e^0:\phi_\mu^0:\phi_\tau^0 = 0:1:0$. In the standard oscillation scenario, these initial ratios result in the detected ratios $\phi_e^\text{std}:\phi_\mu^\text{std}:\phi_\tau^\text{std} \approx 0.22:0.39:0.39$. Alternatively, a pure-$\nu_e$ initial flux, corresponding to $\phi_e^0:\phi_\mu^0:\phi_\tau^0 = 1:0:0$, produced through beta decay has been considered, e.g., in \cite{Lipari:2007su}. In this scenario, high-energy nuclei emmitted by the source have sufficient energy for photodisintegration to occur, but not enough to reach the threshold for pion photoproduction. The neutrons created in the process generate $\overline{\nu}_e$ through beta decay. For these initial ratios, the resulting detected ratios, in the standard oscillation scenario, are $\phi_e^\text{std}:\phi_\mu^\text{std}:\phi_\tau^\text{std} \approx 0.57:0.215:0.215$. The results are summarised in Table \ref{TblStdRS}. In the following sections, we will consider the possibility of observing the hypothetical energy-independent contribution of $H_b$ assuming that the initial ratios correspond to one of these three production scenarios.

\TABLE[t]{
 \begin{tabular}{|l|l|l|l|l|}
  \hline
  Production & Initial flux~ & Std. detected flux & $R^\text{std}$ & $S^\text{std}$ \\
  mechanism & $\phi_e^0:\phi_\mu^0:\phi_\tau^0$ & $\phi_e^\text{std}:\phi_\mu^\text{std}:\phi_\tau^\text{std}$ & &  \\
  \hline
  Pion decay   & $1:2:0~$ & $1:1:1$            & $1$    & $1$ \\
  Muon cooling & $0:1:0~$ & $0.22:0.39:0.39$   & $1.77$ & $1$ \\
  Beta decay   & $1:0:0~$ & $0.57:0.215:0.215$ & $0.38$ & $1$ \\
  \hline
 \end{tabular}
 \caption{Standard values (without energy-independent new physics contributions) of the detected flavour ratios $\phi_\alpha$ ($\alpha = e, \mu, \tau$) and of the ratios-of-ratios $R$, $S$, for the three scenarios of initial flavour ratios considered in the text. The detected ratios were calculated using the average flavour-transition probability in Eq.~(\ref{EqTransProbAvg}) with the central values of the mixing angles: $\sin^2\left(\theta_{12}\right) = 0.304$, $\sin^2\left(\theta_{13}\right) = 0.01$, $\sin^2\left(\theta_{23}\right) = 0.50$. We have defined $R^\text{std} = \phi_\mu^\text{std}/\phi_e^\text{std}$ and $S^\text{std} = \phi_\tau^\text{std}/\phi_\mu^\text{std}$.\label{TblStdRS}}
}

Using the detected flavour ratios, we have defined the ratios of ratios
\begin{equation}
 R = \frac{\phi_\mu}{\phi_e} ~~, ~~~ S = \frac{\phi_\tau}{\phi_\mu} \ .
\end{equation}
Their standard values, $R^\text{std}$ and $S^\text{std}$, i.e., those calculated in the absence of $H_b$, are shown in Table \ref{TblStdRS} for the three choices of initial flavour ratios. Note that $S^\text{std}=1$ for any choice of initial ratios because the value of $\theta_{23}$ used was its best-fit value $\pi/4$, which yields equal detected fluxes of $\nu_\mu$ and $\nu_\tau$ due to maximal mixing. Deviations from this value result in $S \neq 1$ \cite{Xing:2006xd}. In the following sections, when we allow $H_b$ to contribute, we will calculate the extent to which the values of $R$ and $S$ deviate from their standard values.

Since we are considering neutrinos that travel distances of tens or hundreds of Mpc, neutrino decay is a possibility. Flavour mixing and decays in astrophysical neutrinos have been explored before, e.g., in \cite{Maltoni:2008jr,Beacom:2002vi,Barenboim:2003jm} The current strongest direct limit on neutrino lifetime, $\tau/m \gtrsim 10^{-4}$ s eV$^{-1}$, was obtained using solar neutrino data \cite{Beacom:2002cb}. More stringent, though indirect, limits can be obtained by considering neutrino radiative decays and using cosmological data \cite{Mirizzi:2007jd}: $\tau > \text{few} \times 10^{19}$ s or $\tau \gtrsim 5 \times 10^{20}$ s, depending on the mass hierarchy and the absolute mass scale. Assuming that the heaviest mass eigenstates decay into the lightest one plus an undetectable light or massless particle (e.g., a sterile neutrino), then, following \cite{Beacom:2003nh}, the ratio of flavour $\alpha$ at Earth will be
\begin{equation}
  \phi_\alpha 
  = \sum_{\beta=e,\mu,\tau} \sum_i \phi_\beta^0 \lvert \left[U_0\right]_{\beta i}\rvert^2 \lvert \left[U_0\right]_{\alpha i}\rvert^2 e^{-L/\tau_i}
  ~~\underrightarrow{L\gg\tau_i}~~ \sum_{\beta=e,\mu,\tau} \sum_{i\text{(stable)}} \phi_\beta^0 \lvert \left[U_0\right]_{\beta i}\rvert^2 \lvert \left[U_0\right]_{\alpha i}\rvert^2 \ ,
\end{equation}
where $\tau_i$ is the lifetime of the $i$-th mass eigenstate in the laboratory frame. As explained in \cite{Beacom:2003nh}, this expression corresponds to the case where the decay has been completed when the neutrinos arrive at Earth. In a normal hierarchy, $\nu_1$ is the only stable state and so
\begin{equation}
 \phi_\alpha^{\text{dec,norm}} 
 = \lvert \left[U_0\right]_{\alpha1} \rvert^2 \sum_{\beta=e,\mu,\tau} \phi_\beta^0 \lvert \left[U_0\right]_{\beta 1} \rvert^2 \ ,
\end{equation}
while in an inverted hierarchy $\nu_3$ is the stable state and
\begin{equation}
 \phi_\alpha^{\text{dec,inv}} 
 = \lvert \left[U_0\right]_{\alpha3} \rvert^2 \sum_{\beta=e,\mu,\tau} \phi_\beta^0 \lvert \left[U_0\right]_{\beta 3} \rvert^2 \ .
\end{equation}
Reference \cite{Beacom:2003nh} provides expressions for the flavour ratios when the decay product accompanying the lightest eigenstate can also be detected, but since the focus of our analysis is the modification of the oscillations through terms of the form Eq.~(\ref{EqLagrangian}), we have chosen to include only the simplest case of neutrino decay, described by the two preceding expressions.

The reader should bear in mind that there are other mechanisms involving new physics that might also affect the astrophysical flavour fluxes at Earth, such, non-standard interactions \cite{Blennow:2009rp}, coupling of neutrinos to dark energy \cite{Ando:2009ts}, deviations from the unitarity of the PMNS mixing matrix \cite{Xing:2008fg}, to name just a few.


\section{Effect of an energy-indepedent contribution on the astrophysical flavour ratios}\label{Section_Effect}

\subsection{Detection of $\nu_\mu$}\label{Section_Effect_Sub_OnlyNuMu}

\FIGURE[t!]{
 \scalebox{0.6}{\includegraphics{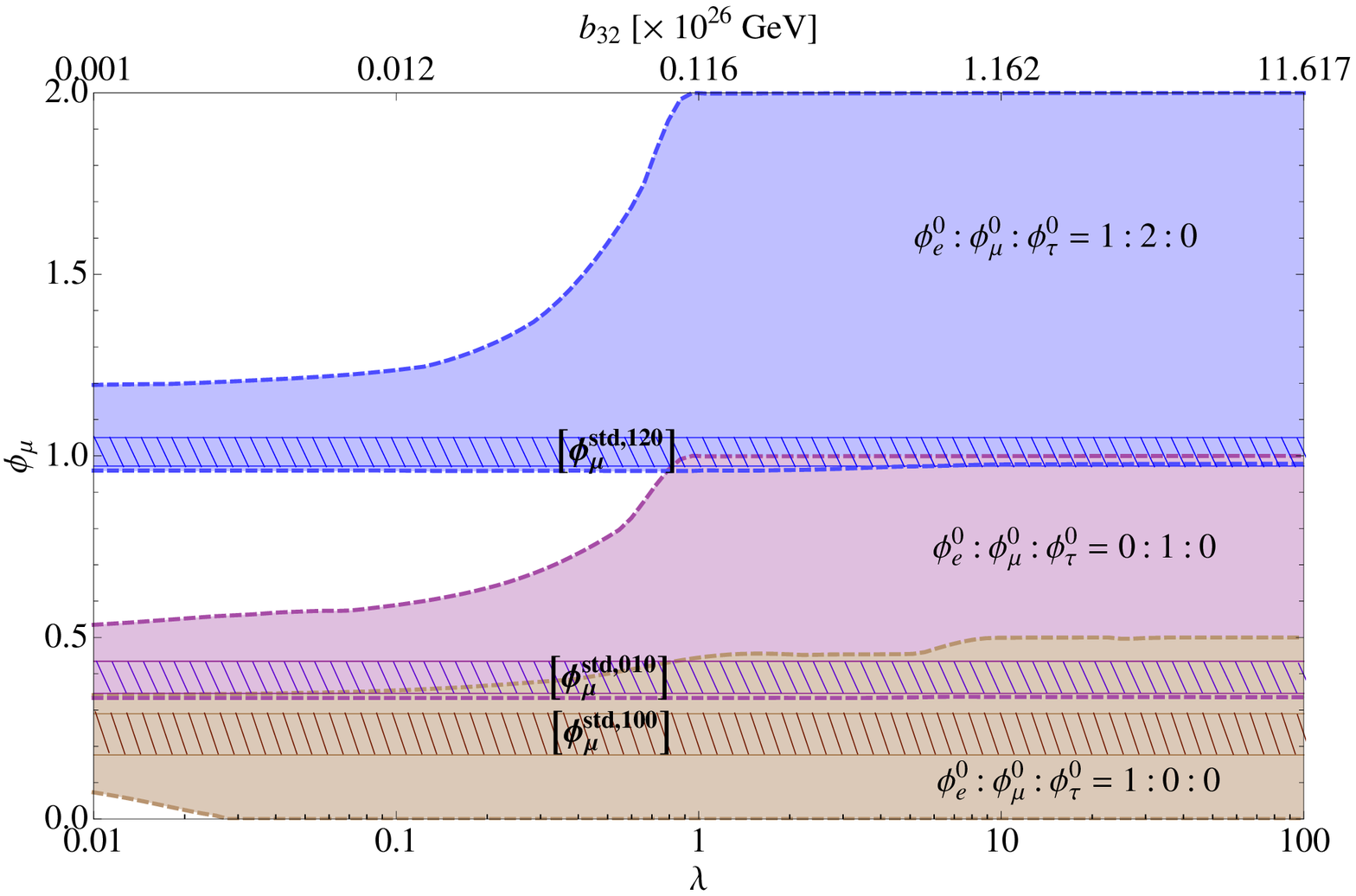}} 
 \caption{Allowed regions of values of the detected muon-neutrino flavour ratio, $\phi_\mu$, as a function of $\lambda$ for different neutrino production models. The three standard mixing angles ($\theta_{12}$, $\theta_{13}$, $\theta_{23}$) were varied within $3\sigma$ bounds and the three CPTV angles ($\theta_{b12}$, $\theta_{b13}$, $\theta_{b23}$) were varied within $\left[0,\pi\right]$, while $\delta_\text{CP} = 0$. The hatched regions are the allowed regions of $\phi_\mu$ when only standard oscillations are allowed, and allowing the $\theta_{ij}$ to vary within their $3\sigma$ bounds.}
 \label{Fig_OnlyPhimu}
}
  
In Figure \ref{Fig_OnlyPhimu} we present a plot of the muon-neutrino flavour ratio, $\phi_\mu$, as a function of $\lambda$. The coloured bands correspond to different neutrino production scenarios, namely: $\phi_e^0:\phi_\mu^0:\phi_\tau^0 = 1:2:0$ (blue), $0:1:0$ (purple) and $1:0:0$ (brown), which have been generated by varying the three CPTV angles ($\theta_{b12}$, $\theta_{b13}$, $\theta_{b23}$) within $\left[0,\pi\right]$, and the three standard mixing angles ($\theta_{12}$, $\theta_{13}$, $\theta_{23}$) within their $3\sigma$ bounds, with $\delta_\text{CP} = 0$. For comparison, we have included the hatched bands which represent the pure standard oscillation case, that is, without CPTV. These have been generated by setting $\lambda = 0$ and varying the three standard mixing angles within their $3\sigma$ bounds, again with $\delta_\text{CP} = 0$. As expected, the standard-oscillation bands are contained within the corresponding CPTV region.

When CPTV is allowed, we observe large deviations of $\phi_\mu$ with respect to the pure standard-oscillation bands, especially for the scenarios $\phi_e^0:\phi_\mu^0:\phi_\tau^0 = 1:2:0$ and $0:1:0$, and less so for the scenario $1:0:0$. Starting from $\lambda \sim 0.1$ ($b_{32} \simeq 1.2\times10^{-28}$ GeV), the CPTV bands start growing with $\lambda$, as was expected, since $\lambda$ measures the strength of the CPT violation. Thus, the influence of the CPTV contribution to the oscillations grows and, as a consequence, the accessible region also grows. This is due to the wide range of values that the CPTV mixing angles can take, in comparison with the standard ones. Past $\lambda = 1$, the CPTV regions reach a plateau, owing to the fact that the CPTV term becomes dominant over the standard-oscillation term in the Hamiltonian.

An interesting feature is the overlap between the standard-oscillation band for the scenario $1:2:0$ and the CPTV region for scenario $0:1:0$. A similar overlap occurs between the scenarios $0:1:0$ (without CPTV) and $1:0:0$ (with CPTV). As a consequence of these overlaps, if CPTV exists for certain values of the parameters, a measurement of $\phi_\mu$ will be insufficient to distinguish what the neutrino production model is. For instance, if a value of $\phi_\mu \simeq 0.4$ were measured, and $\lambda \gtrsim 0.2$, we would not be able to assert whether the initial fluxes were $0:1:0$ or $1:0:0$. Analogously, since the standard-oscillation bands are contained within the CPTV regions, for these cases we will be unable to conclude, from the measurement of $\phi_\mu$, whether or not CPTV effects are present. 

Although we have not presented it here, we have tested that when varying $\delta_\text{CP}$ within $\left[0,2\pi\right]$, the regions change only in a few percent. Therefore, the features observed in Fig.~\ref{Fig_OnlyPhimu} are largely independent of the value of $\delta_\text{CP}$.
     
\subsection{Detection of two flavours: $\nu_\mu$ and $\nu_e$}\label{Section_Effect_Sub_NuMu_NuE}

In this section, we consider an scenario where only muon- and electron-neutrinos are detected in an available neutrino telescope. Assuming that flavour identification is possible, we can define the ratio of observed flavour ratios
\begin{equation}
 R \equiv \phi_\mu / \phi_e \ ,
\end{equation}
which, in the presence of CPTV, depends on $\lambda$ and on the three CPTV mixing angles through the definition of the $\phi_\alpha$ in terms of the flavour-transition probabilities. Figure \ref{Fig_RvsLambda} shows the regions of values of $R$ as a function of $\lambda$ by varying the three CPTV angles within $\left[0,\pi\right]$.

\FIGURE[b!]{
 \scalebox{0.6}{\includegraphics{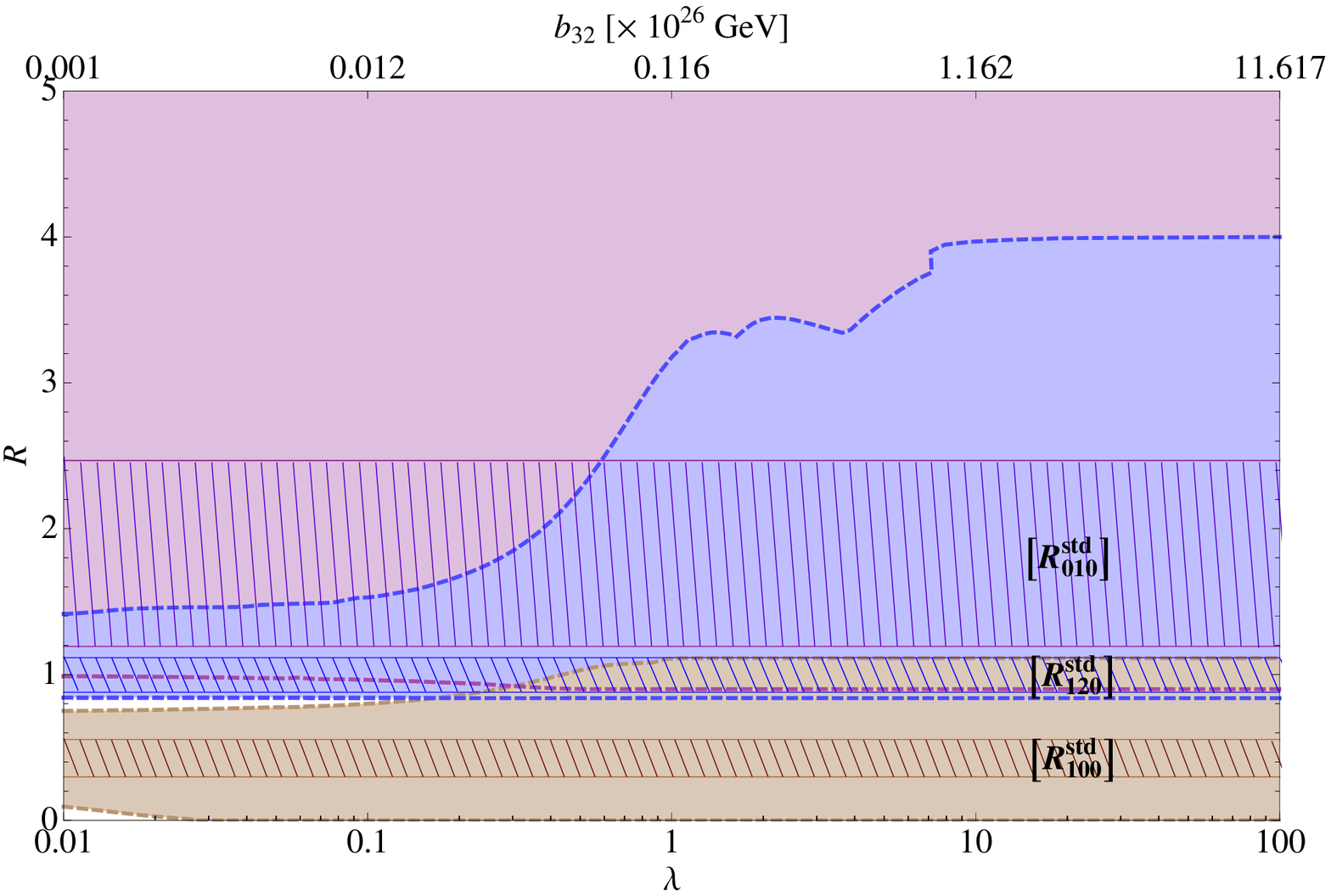}} 
 \caption{Allowed regions of values of $R \equiv \phi_\mu/\phi_e$ as a function of $\lambda$, for three scenarios of initial flavour ratios: $\phi_e^0:\phi_\mu^0:\phi_\tau^0 = 1:2:0$ (blue), $0:1:0$ (purple) and $1:0:0$ (brown). For each one of them, at each value of $\lambda$, the three CPTV mixing angles, $\theta_{b12}$, $\theta_{b13}$ and $\theta_{b23}$, were independently varied within $\left[0,\pi\right]$, and the three standard mixing angles, $\theta_{12}$, $\theta_{13}$ and $\theta_{23}$, were varied within their $3\sigma$ bounds (Eq.~(\ref{EqStdMixPar})): the lowest and highest value of $R$ obtained in this way define, respectively, the lower and upper bounds of the corresponding region, for this particular value of $\lambda$. The CP-violating phase $\delta_\text{CP} = 0$. The upper horizontal axis is $b_{32} = \lambda ~\Delta m_{32}^2 / \left(2 E^\ast\right)$, with $E^\ast = 1$ PeV; to find the corresponding values of $b_{21}$, notice that $b_{21}/b_{32} = \Delta m_{21}^2 / \Delta m_{32}^2 \simeq 1/30$. The hatched horizontal bands are the allowed regions of $R$ assuming only standard oscillations and allowing the standard mixing angles $\theta_{ij}$ to vary within their $3\sigma$ bounds.}
 \label{Fig_RvsLambda}
}

The coloured regions in Fig.~\ref{Fig_RvsLambda} grow with $\lambda$, for the same reason as they did in Fig.~\ref{Fig_OnlyPhimu}. Under the assumption of a $0:1:0$ production model, $R$ can attain very large values, between $10^6$ and $10^7$, as a result of very low electron-neutrino fluxes. On the other hand, the CPTV region associated to the $1:2:0$ production model reaches a maximum of $R = 2$ after $\lambda \simeq 5$, while the one associated to $1:0:0$ reaches a maximum of $R = 1$ after $\lambda \simeq 1$. Therefore, a measurement of $R \gg 4$ could imply that the production model is $0:1:0$, and that CPTV effects are present, but will not be enough to set strong bounds on $\lambda$. The minimum value in both the $0:1:0$ and $1:2:0$ regions is located around $R=0.8\sim1$, while for the $1:0:0$ flux, it is zero for most of the range of $\lambda$. 

If a value of $R \lesssim 4$ is found, the ability to single out a production model depends on the exact value of $R$ that is measured. A single production model can be distinguished univocally for some measured values of $R$ (e.g., for $R < 0.84$, the model is $1:0:0$), while for others, two (when $1 \lesssim R \lesssim 4$, depending on $\lambda$) or three models (when $R \simeq 1$) can account for the same measured value. In the same way, for some values of $R$, CPTV and standard oscillations cannot be distinguished. The conclusions, from Fig.~\ref{Fig_RvsLambda}, that can be obtained from the measurement of different values of $R$ are shown in Table \ref{Tbl_OnlyR}.

\TABLE[t!]{
 \begin{tabular}{|c|l|}
  \hline
  Measured $R$ & Conclusion \\
  \hline
  $R > 4$                                        & Initial ratios $0:1:0$ and CPTV \\
  $\left[R_{010}^{\text{std}}\right] < R < 4$    & Initial ratios $0:1:0$ or $1:2:0$, and CPTV \\
  $R \in \left[R_{010}^{\text{std}}\right]$      & (Initial ratios $0:1:0$ and std. osc.) or ($1:2:0$ and CPTV) \\
  $\left[R_{120}^{\text{std}}\right] < R < \left[R_{010}^{\text{std}}\right]$ 
                                                 & Initial ratios $0:1:0$ or $1:2:0$, and CPTV \\
  $R \in \left[R_{120}^{\text{std}}\right]$      & (Initial ratios $1:2:0$ and std. osc.) or \\
                                                 & (initial ratios $0:1:0$ or $1:0:0$, and CPTV) \\
  $0.84 < R < \left[R_{120}^{\text{std}}\right]$ & Initial ratios $1:2:0$ or $1:0:0$, and CPTV \\
  $\left[R_{100}^{\text{std}}\right] < R < 0.84$ & Initial ratios $1:0:0$ and CPTV \\
  $R \in \left[R_{100}^{\text{std}}\right]$      & Initial ratios $1:0:0$ and std. osc. \\
  $R < \left[R_{100}^{\text{std}}\right]$        & Initial ratios $1:0:0$ and CPTV \\
  \hline
 \end{tabular}
 \caption{Conclusions that can be obtained depending on the measured value of $R$, according to Fig.~\ref{Fig_RvsLambda}. $\left[R_{120}^{\text{std}}\right]$, $\left[R_{010}^{\text{std}}\right]$ and $\left[R_{100}^{\text{std}}\right]$ represent, respectively, the standard-oscillation bands corresponding to the $\phi_e^0:\phi_\mu^0:\phi_\tau^0 = 1:2:0$, $0:1:0$ and $1:0:0$ production models. When we refer to standard oscillations, we mean either the inexistence of CPTV ($\lambda = 0$) or the existence of too small a CPTV ($\lambda \ll 1$).\label{Tbl_OnlyR}}
}

\subsection{Detection of three flavours: $\nu_\mu$, $\nu_e$ and $\nu_\tau$}\label{Section_Effect_Sub_NuMu_NuE_NuTau}

\EPSFIGURE[b!]{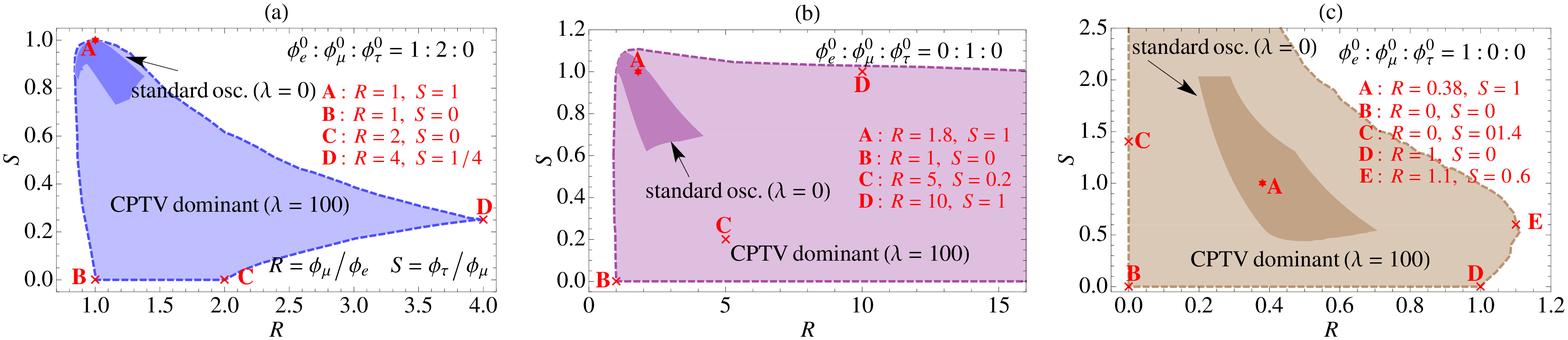,width=15.5cm} 
 {Regions of $R$ and $S$ accessible with CPTV by assuming different neutrino production scenarios. Different colours correspond to different initial ratios: $\phi_e^0:\phi_\mu^0:\phi_\tau^0 = 1:2:0$ (blue), $0:1:0$ (purple), $1:0:0$ (brown). Darker regions are generated with standard (CPT-conserving) neutrino flavour oscillations, by allowing the standard mixing angles to vary within their $3\sigma$ experimental bounds. Lighter regions correspond to the case when we include a dominant CPTV contribution with $\lambda= 100$, and allow the CPTV angles $\theta_{bij}$ to vary within $\left[0,\pi\right]$. All phases are set to zero.\label{FigScatterSR_notablepts}}

Given the approach of this work, we have considered as a natural step to extend our analysis by taking into account the possibility of tau-neutrino detection. For this purpose, we have defined the ratio
\begin{equation}
 S = \phi_\tau/\phi_\mu \ ,
\end{equation}
and studied the effects of the new physics in the $R$ vs.~$S$ plane. 

Before we describe the results of this section, it will be useful to turn our attention to the values of $R$ and $S$ associated to the different production models when only standard oscillations are allowed, along with the correspondong standard values of the detected fluxes $\phi_\alpha$. These values are shown in Table \ref{TblStdRS}. The CPTV regions in Figures \ref{FigScatterSR_notablepts}\---\ref{FigScatterPlots_VarLambda} have been generated by setting $\lambda = 100$ (Figs.~\ref{FigScatterSR_notablepts} and \ref{FigScatterSR_3sigma}) and $\lambda = 1, 10, 100$ (Fig.~\ref{FigScatterPlots_VarLambda}), fixing the standard mixing parameters at their best-fit values (see Eq.~(\ref{EqStdMixPar})), and varying the CPTV mixing angles $\theta_{bij}$ within the range $\left[0,\pi\right]$, while the standard-oscillation regions have been generated by setting $\lambda = 0$ and varying the standard mixing angles within their $3\sigma$ bounds. With the exception of Figure \ref{Fig_StdOscRSDependenceOnDeltaCP}, where $\delta_\text{CP}$ has been allowed to vary, we have set all phases equal to zero.

In Fig.~\ref{FigScatterSR_notablepts} we display in three $R$ vs.~$S$ panels the allowed regions of values that correspond to pure standard oscillations at $3\sigma$, in dark tones, and the corresponding regions allowed by the mixed solution composed of the standard oscillation plus CPTV effects, with $\lambda = 100$, in lighter tones. At this value of $\lambda$, the CPTV contributions are the dominant ones in the Hamiltonian, i.e., $H_f \simeq H_b$. Each plot corresponds to a different neutrino production scenario: (a) to $1:2:0$, (b) to $0:1:0$, and (c) to $1:0:0$. 

From these plots we can extract two observations: one is the potentially dramatic deviation of the allowed values of the pair $\left(R,S\right)$ when CPTV is turned on, and the other is the presence of points that are common to all scenarios. The latter implies that, if there is CPTV, there could exist pairs $\left(R,S\right)$, such as $\left(1,0\right)$, that can be generated by any of the three production models, each with a different set of values for the CPTV mixing angles. There are also $\left(R,S\right)$ pairs which could be generated by one production model with standard oscillations or with a different model with CPTV, e.g., those lying around $\left(1,0.95\right)$. It is convenient to remark that this figure and Fig.~\ref{Fig_RvsLambda} are consistent with each other, which can be shown by projecting the CPTV regions of Fig.~\ref{FigScatterSR_notablepts} onto the horizontal axis and checking that the limits on $R$ agree with those on Fig.~\ref{Fig_RvsLambda}. We have marked a few notable points in each plot: for the three production models, the points labelled with A correspond to the best-fit values of the standard-oscillation mixing parameters in Eq.~(\ref{EqStdMixPar}).

\FIGURE[t!]{
 \scalebox{0.6}{\includegraphics{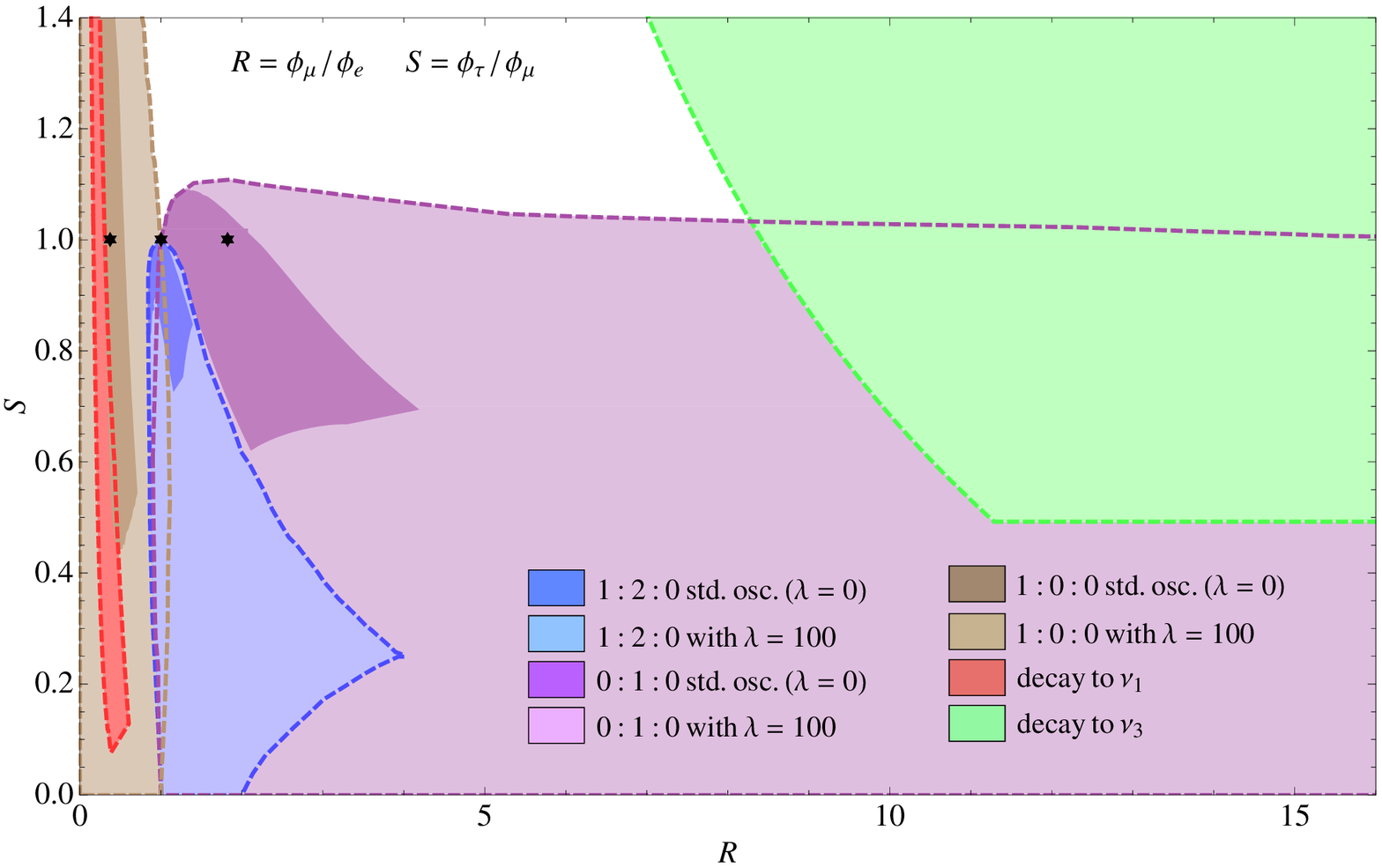}} 
 \caption{Regions of $R$ and $S$ accessible by assuming different neutrino production scenarios: $\phi_e^0:\phi_\mu^0:\phi_\tau^0 = 1:2:0$ (blue), $0:1:0$ (purple) and $1:0:0$ (brown). Darker shades of blue, purple and brown correspond to standard, CPT-conserving (i.e., $\lambda =0$), flavour transitions, while lighter shades correspond to flavour transitions dominated by CPT violation ($\lambda = 100$). The former were generated by varying the standard mixing angles $\theta_{ij}$ within their current $3\sigma$ experimental bounds; the latter, by fixing the $\theta_{ij}$ to their best-fit values and varying the new mixing angles $\theta_{bij}$ within $0$ and $\pi$. All of the phases were set to zero: $\delta_\text{CP} = \delta_b = \phi_{b2} = \phi_{b3} =0$. Also shown are the regions accessible through decay of the neutrinos into invisible products, when $\nu_1$ is the lightest mass eigenstate (red) and when $\nu_3$ is the lightest one.}
 \label{FigScatterSR_3sigma}
}

In Fig.~\ref{FigScatterSR_3sigma}, the allowed $R\--S$ regions corresponding to the three neutrino production models are shown together: $1:2:0$ in blue, $0:1:0$ in purple, and $1:0:0$ in brown, where, as before, the darker tones correspond to pure standard oscillations, and the lighter tones, to a dominant CPTV with $\lambda = 100$. Following the argument in Section \ref{Section_Theory_Sub_FlavRatios}, we have also included the regions of $\left(R,S\right)$ pairs allowed by neutrino decay into invisible products, considering both the cases of a normal mass hierarchy (decay into the $\nu_1$ eigenstate), in red, and of an inverted hierarchy (decay into $\nu_3$), in green. The decay regions were also generated by varying the standard-oscillation mixing angles within their $3\sigma$ bounds.

While the regions corresponding to $1:2:0$ do not overlap those corresponding to neutrino decay, this is not the case for the $0:1:0$ scenario, for which there is a clear overlap between the CPTV region and the one from neutrino decay into $\nu_3$. The region corresponding to decay into $\nu_1$ is completely contained within the $1:0:0$ CPTV region. Projecting onto the horizontal axis, we see that, apart from the same overlaps between the allowed $R$ intervals that existed in the two-neutrino case, as observed in Fig.~\ref{Fig_RvsLambda}, new overlaps appear due to the inclusion of the neutrino-decay regions. Particularly, whereas prior to the inclusion of decays, a value of $R \gg 4$ was a clear indication of the existence of CPTV and of a $0:1:0$ production model, now Fig.~\ref{FigScatterSR_3sigma} shows that, after $R = 7$, the decay into $\nu_3$, assuming an inverted mass hierarchy, can also be accountable for a high value of $R$. On the other hand, values of $0.25 \lesssim R \lesssim 0.75$ can be reached by the $1:0:0$ production model, with or without CPTV, as well as with neutrino decay into $\nu_1$, assuming a normal hierarchy. A normal mass hierarchy is able to yield values of $S$ as low as $\sim 0.075$, leaving only the small window between zero and this value as unique feature of CPTV. Taking into account experimental uncertainty, it is likely that this window is actually non-existent. Only if the measured $S \gtrsim 1.1$, and in the absence of decays, would it be possible to identify a single production model, $1:0:0$, as the one responsible. If decays are allowed, however, they can also account, irrespectively of the mass hierarchy, for $S > 1.1$, and the ability to single out a production model is lost. Thus, the signatures of CPTV become less unique in the presence of decays. A more complete analysis of neutrino decays \cite{Barenboim:2003jm}, exploring also the possibility of incomplete decays and decay into visible products, further reduces our ability to uniquely identify the presence of CPTV.

In the absence of neutrino decays, a measurement of $R \gtrsim 4.1$ and $S \lesssim 1.1$ could indicate the presence of dominant CPTV and a $0:1:0$ production model. Also, for $R \lesssim 0.9$, the production model is $1:0:0$; if $S \lesssim 0.45$, there is dominant CPTV and, for other values of $S$, the existence of CPTV will depend on the value of $R$ measured. Close to $R = 1$, and for $S \lesssim 0.9$, any of the three production models with dominant CPTV can explain the measured $\left(R,S\right)$ pair, while for $0.9 \lesssim S < 1$, the pair could be generated either by the $0:1:0$ or $1:0:0$ models with CPTV, or by the $1:2:0$ model with standard oscillations. For $1.1 \lesssim R < 4$ and $S \lesssim0.6$, there is a large region of overlap between the $1:2:0$ and $0:1:0$ models with dominant CPTV. 

\FIGURE[t!]{
 \scalebox{0.6}{\includegraphics{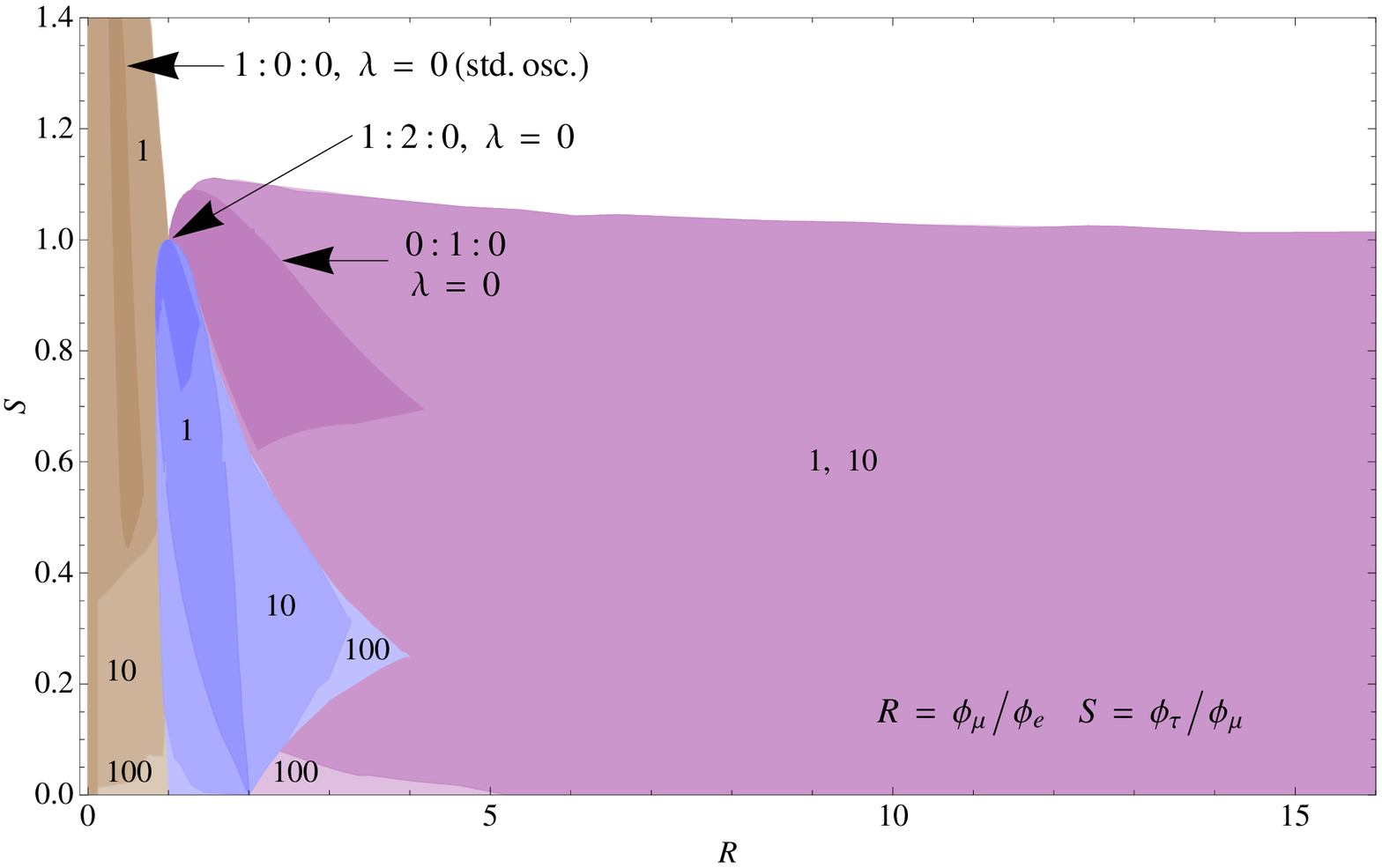}} 
 \caption{Regions of values of $R$ and $S$ accessible when varying the parameter $\lambda$ between 0 (no CPT breaking) and 100 (dominant CPTV term), for different neutrino production models. In blue: $\phi_e^0:\phi_\mu^0:\phi_\tau^0 = 1:2:0$; in purple, $0:1:0$; in brown, $1:0:0$. The region corresponding to pure standard oscillations ($\lambda = 0$) was generated by varying the standard mixing angles within their $3\sigma$ experimental bounds, Eq.~(\ref{EqStdMixPar}). To generate the regions corresponding to $\lambda = 1, 10, 100$, the standard mixing angles were fixed to their best-fit values, and the three CPTV mixing angles $\theta_{bij}$ were varied independently within $\left[0,\pi\right]$. The CP-violating phase $\delta_\text{CP} = 0$ for all of the regions.}
 \label{FigScatterPlots_VarLambda}
}

Fig.~\ref{FigScatterPlots_VarLambda} shows the effect of the variation of the parameter $\lambda$ between 0 (no CPTV, i.e., $H_f = H_m$) and 100 (dominant CPTV term, i.e., $H_f \simeq H_b$) on the allowed $R \-- S$ regions, for the different neutrino production scenarios. For the three of them, we can observe significant deviations from the predictions of the standard-oscillation case, even for low values of $\lambda$. In this sense, it is interesting to point out that in the case of an experimental non-detection of CPTV in the neutrino flavour ratios, $R$ and $S$ can be used to set limits on the related parameters. In fact, when $\lambda = 1$, and for a neutrino energy of 1 PeV, we can attain limits for the CPTV eigenvalues $b_{ij}$ in the order of $10^{-29}$ and $10^{-27}$ GeV, for $b_{21}$ and $b_{23}$, respectively. It is also important to mention that these results can be easily rescaled to any energy, just by doing $b_{ij} \times$(PeV/$E$). This would mean a very significant improvement over the current bounds of $10^{-23} \-- 10^{-21}$ GeV for $b_{21}$ and $b_{32}$, respectively \cite{Dighe:2008bu}.

Finally, it is worth exploring the effect of the choice of value of $\delta_\text{CP}$ on $R$ and $S$. From Fig.~\ref{Fig_StdOscRSDependenceOnDeltaCP}, we see that the effect on $R$ of a non-zero value of $\delta_\text{CP}$ in the standard-oscillation regions is more prominent for a choice of initial ratios of $0:1:0$, and less so for $1:2:0$ and $1:0:0$. For $S$, the effect of a non-zero phase is greater for $1:0:0$, and less for $0:1:0$ and $1:2:0$. Note that using a non-zero value of $\delta_\text{CP}$ can lead to a variation in $R$ of up to $42\%$ (in the $0:1:0$ case) and in $S$ of up to $40\%$ (in the $1:0:0$ case) with respect to the standard case of $\theta_{13}=\delta_\text{CP}=0$, when the largest $3\sigma$ allowed value of $\sin^2\left(\theta_{13}\right) = 0.056$ is assumed.

\EPSFIGURE[t!]{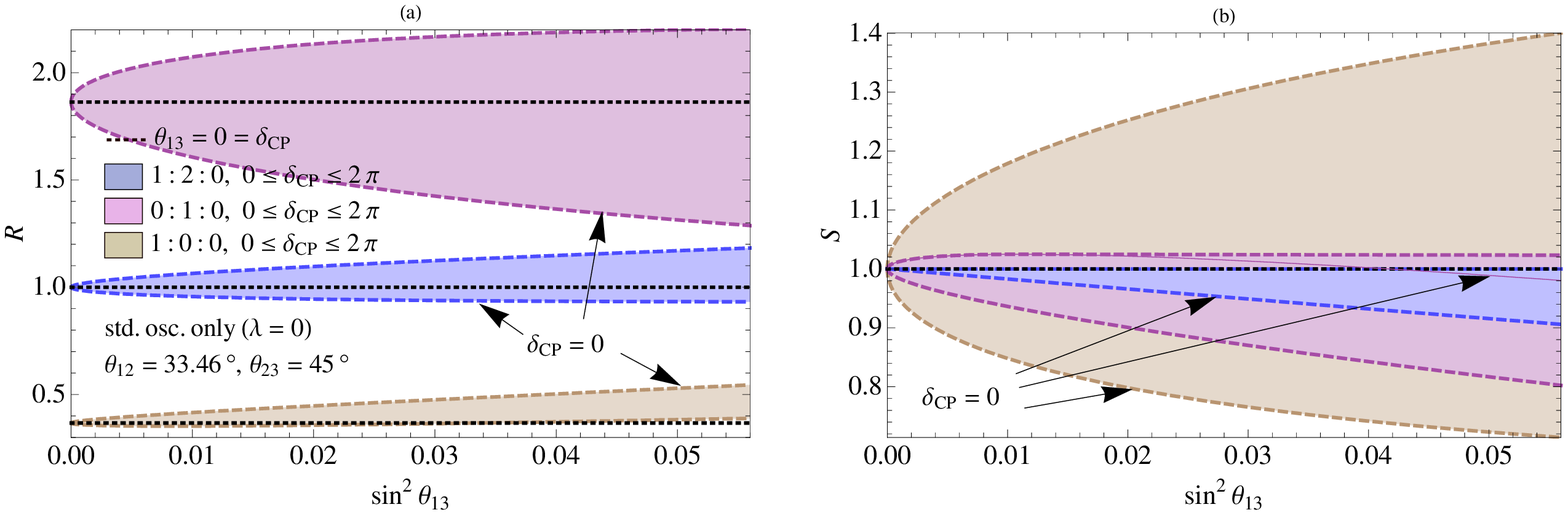,width=15.5cm}{Variation of $R$ and $S$ with $\theta_{13}$, when the CP-violation phase $\delta_\text{CP}$ is allowed to vary between $0$ and $2\pi$. The upper limit for $\theta_{13}$ is given by the current bound $\sin^2\left(\theta_{13}\right) \le 0.056$ ($3\sigma$) and the mixing parameters $\theta_{12}$, $\theta_{23}$, $\Delta m_{21}^2$ and $\Delta m_{32}^2$ have been set to their current best-fit values; standard flavour oscillations have been assumed throughout (i.e., $\lambda=0$). Three different scenarios of initial flavour ratios have been considered: $\phi_e^0:\phi_\mu^0:\phi_\tau^0 = 1:2:0$ (in blue), $0:1:0$ (in purple) and $1:0:0$ (in brown). To allow for comparison, the arrows point to the curves on which $\delta_\text{CP} = 0$.\label{Fig_StdOscRSDependenceOnDeltaCP}}

\section{Detection prospects at IceCube considering CPTV}\label{Section_IceCube}

\subsection{Experimental setup}\label{Exp_setup}

The IceCube neutrino telescope \cite{DeYoung:2009er,Halzen:2009xm}, located at the South Pole, can detect both muons and showers initiated by incoming high-energy astrophysical neutrinos. Muon-neutrinos are detected through the high-energy muon produced in charged-current (CC) deep inelastic neutrino-nucleon scattering: the \u{C}erenkov light emitted by the fast-moving muon in ice is detected by the photomultipliers buried in ice and used to reconstruct the muon track. Electron-neutrinos can generate electromagnetic and hadronic showers in CC interactions. Tau-neutrino CC events have a distinct topology: they either show two hadronic showers joined by a tau track (a double bang) or a tau track ending with the decay of the tau in a hadronic shower (lollipop). Additionally, every flavour of neutrino is able to generate hadronic showers in neutral-current (NC) interactions. Flavour identification is expected to be difficult at IceCube \cite{Beacom:2003nh} and the number of lollipops expected is about one every two years (for a neutrino flux of $10^{-7}$ GeV$^{-1}$ s$^{-1}$ cm$^{-2}$ sr$^{-1}$ \cite{Beacom:2003nh}), so the useful observables turn out to be the total number of NC plus CC showers, $N_\text{sh} = N_\text{sh}^\text{NC} + N_\text{sh}^\text{CC}$ and the number of muon tracks, $N_{\nu_\mu}$. From this, we can construct the closest experimental analogue of the variable $R \equiv \phi_\mu / \phi_e$ as $R_\text{exp} \equiv N_{\nu_\mu}/N_\text{sh}$. Due to the very low number of tau-neutrinos expected at IceCube, there is no practical experimental analogue of $S \equiv \phi_\tau/\phi_\mu$.

Following our analysis of CPTV in the previous sections, where we set the scale of CPTV at $E^\ast = 1$ PeV, we have adopted the energy range $E_\nu^{\min} = 10^6 \le E_\nu/\text{GeV} \le E_\nu^{\max} = 10^{12}$ for our predictions. Within this energy range, the Earth is opaque to neutrinos due to the increased number of NC interactions which degrade their energy \cite{Gandhi:1995tf}, so we have calculated only the number of downgoing events. Bear in mind, however, that due to the tighter background filtering that is required for the observation of downgoing neutrinos, which we have not taken into account in the following calculations, our estimates may be optimistic. Conveniently, for most of this range the atmospheric $\nu_\mu$ background flux will be well below the fluxes from the neutrino production models that we have probed (see Figure \ref{FigFluxes}). Using the expressions of \cite{Anchordoqui:2005gj}, we can estimate the number of CC and NC events at IceCube, for a given astrophysical all-flavour diffuse neutrino flux, denoted here by $\Phi^{\nu_\text{all}}$:
\begin{eqnarray}
 \label{EqNEvtsCC}
 N_{\nu_\text{all}}^\text{CC} &=& T n_T V_\text{eff} \Omega 
                                  \int_{E_\text{sh}^{\min}}^{E_\nu^{\max}} \Phi^{\nu_\text{all}}\left(E_\nu\right) \frac{1}{2}
				  \left[ \sigma_\text{CC}^{\nu N}\left(E_\nu\right) + \sigma_\text{CC}^{\overline{\nu} N}\left(E_\nu\right) \right] ~dE_\nu \\
 \label{EqNEvtsNC1}
 N_{\nu_\text{all}}^\text{NC} &=& T n_T V_\text{eff} \Omega 
			          \int_{E_\text{sh}^{\min}}^{E_\nu^{\max}} ~dE_\nu
				  \int_{E_\nu-E_\text{sh}^{\min}}^{E_\nu^{\max}} ~dE_\nu^\prime
				  \Phi^{\nu_\text{all}}\left(E_\nu\right) \times \nonumber \\ &&
				  \times \frac{1}{2}
				  \left[ \frac{d\sigma_\text{NC}^{\nu N}}{dE_\nu^\prime}\left(E_\nu,E_\nu^\prime\right)
				  + \frac{d\sigma_\text{NC}^{\overline{\nu} N}}{dE_\nu^\prime}\left(E_\nu,E_\nu^\prime\right) \right] \ ,
\end{eqnarray}
where $E_\nu^\prime$ is the energy of the secondary neutrino in the NC interaction, $T$ is the exposure time, $n_T = 5.1557 \times 10^{23}$ cm$^{-3}$ is the number density of targets (nucleons) in ice, $V_\text{eff}$ is the effective detector volume (1 km$^3$ for IceCube) and $\Omega = 5.736$ sr is the detector's opening angle (up to 85$^\circ$). The total cross sections for neutrino and anti-neutrino deep inelastic scattering, $\sigma_\text{CC}^{\nu N}$, $\sigma_\text{CC}^{\overline{\nu} N}$, $\sigma_\text{NC}^{\nu N}$ and $\sigma_\text{NC}^{\overline{\nu} N}$, have been extracted from \cite{Gandhi:1998ri}. The differential cross sections used are written as a function of both the primary and the secondary neutrino energies, which in NC interactions are related through $E_\nu^\prime = \left( 1 - \left\langle y_\text{NC} \right\rangle \right) E_\nu$, with $\left\langle y_\text{NC} \right\rangle$ the NC inelasticity parameter, extracted from \cite{Gandhi:1995tf}. If the interval of interest ($10^6 \-- 10^{12}$ GeV) is partitioned into small enough subintervals, then within each, the NC cross section can be approximated by $\sigma_\text{NC} = A E_\nu^B$ with $A$, $B$ constants for each subinterval. We can then write
\begin{equation}
 \sigma_\text{NC}^{\nu N} = A E_\nu^B = A \left[ \left( 1 - \left\langle y_\text{NC} \left(E_\nu\right) \right\rangle \right)^{-1} E_\nu^\prime\right]^B \ ,
\end{equation}
so that
\begin{equation}
 \frac{d\sigma_\text{NC}}{dE_\nu^\prime}\left(E_\nu,E_\nu^\prime\right)
 = A B \left[ 1 - \left\langle y_\text{NC} \left(E_\nu\right) \right\rangle \right]^{-B} \left(E_\nu^\prime\right)^{B-1} \ .
\end{equation}
Using this expression in Eq.~(\ref{EqNEvtsNC1}) and performing the $E_\nu^\prime$ integral, we obtain the simplified form
\begin{eqnarray}\label{EqNEvtsNC2}
 N_{\nu_\text{all}}^\text{NC} \simeq &\frac{1}{2}& T n_T V_\text{eff} \Omega \int_{E_\text{sh}^{\min}}^{E_\nu^{\max}} ~dE_\nu
                                     \Phi^{\nu_\text{all}}\left(E_\nu\right) \times \nonumber \\
                                     &\times& \left[
                                     \frac{\sigma_\text{NC}^{\nu N}\left(E_\nu^{\max}\right) - \sigma_\text{NC}^{\nu N}\left(E_\nu-E_\text{sh}^{\min}\right)}{\left[ 1 - \left\langle y_\text{NC}^{\nu N}\left(E_\nu\right) \right\rangle \right]^B} +
                                     \frac{\sigma_\text{NC}^{\overline{\nu} N}\left(E_\nu^{\max}\right) - \sigma_\text{NC}^{\overline{\nu} N}\left(E_\nu-E_\text{sh}^{\min}\right)}{\left[ 1 - \left\langle y_\text{NC}^{\overline{\nu} N}\left(E_\nu\right) \right\rangle \right]^{\overline{B}}} 
                                     \right] \ ,
\end{eqnarray}
with $B$ and $\overline{B}$ taking the appropriate values in each subinterval.

The number of CC showers generated by electron- and tau-neutrinos are, respectively, $N_{\text{sh},e}^\text{CC} = \tilde{\phi}_e N_{\nu_\text{all}}^\text{CC}$ and $N_{\text{sh},\tau}^\text{CC} = \tilde{\phi}_\tau N_{\nu_\text{all}}^\text{CC}$, where
\begin{equation}\label{NormPhi}
 \tilde{\phi}_\alpha \equiv \frac{\phi_\alpha}{\phi_e+\phi_\mu+\phi_\tau} \in \left[0,1\right]
\end{equation}
are the normalised flavour ratios. The number of CC showers is given by
\begin{equation}
 N_\text{sh}^\text{CC} = N_{\text{sh},e}^\text{CC} + N_{\text{sh},\tau}^\text{CC} = \left( \tilde{\phi}_e + \tilde{\phi}_\tau \right) N_{\nu_\text{all}}^\text{CC} 
\end{equation}
and the total number of CC plus NC showers is therefore
\begin{equation}\label{EqNumTotSh}
 N_\text{sh} = N_\text{sh}^\text{CC} + N_\text{sh}^\text{NC} = \left( 1 - \tilde{\phi}_\mu \right) N_{\nu_\text{all}}^\text{CC} + N_{\nu_\text{all}}^\text{NC} \ ,
\end{equation}
where we have used the fact that $\sum_{\beta=e,\mu,\tau} \tilde{\phi}_\beta = 1$. In a similar way, the number of downgoing muon-neutrinos is given by
\begin{equation}\label{EqNumNuMu}
 N_{\nu_\mu} = \tilde{\phi}_\mu N_{\nu_\text{all}}^\text{CC} \ .
\end{equation}
Using Eqs.~(\ref{EqNEvtsCC}), (\ref{EqNEvtsNC2}), (\ref{EqNumTotSh}) and (\ref{EqNumNuMu}), we find
\begin{equation}\label{EqRexp}
 R_\text{exp} 
 = \frac{N_{\nu_\mu}}{N_\text{sh}}
 = \frac{\tilde{\phi}_\mu}{\left(1-\tilde{\phi}_\mu\right)+N_{\nu_\text{all}}^\text{NC}/N_{\nu_\text{all}}^\text{CC}} ~,
\end{equation}
and we see that $R_\text{exp}$ depends on only one of the normalised flavour ratios, $\tilde{\phi}_\mu$, and that it is independent of the exposure time and the effective detector size. Increasing $T$ and $V_\text{eff}$, however, results in a larger event yield and consequently in lower statistical uncertainty. Considering $N_{\nu_\text{all}}^\text{NC}$ and $N_{\nu_\text{all}}^\text{CC}$ as independent variables with Poissonian errors, i.e., $\sqrt{N_{\nu_\text{all}}^\text{NC}}$ and $\sqrt{N_{\nu_\text{all}}^\text{CC}}$ respectively, we find the error on $R_\text{exp}$ to be
\begin{equation}\label{EqSigmaRexp}
 \sigma_{R_\text{exp}} = \frac{R_\text{exp}}{\left(1-\tilde{\phi}_\mu\right)+N_{\nu_\text{all}}^\text{NC}/N_{\nu_\text{all}}^\text{CC}} 
                         \frac{N_{\nu_\text{all}}^\text{NC}}{N_{\nu_\text{all}}^\text{CC}}
                         \sqrt{\frac{1}{N_{\nu_\text{all}}^\text{NC}} + \frac{1}{N_{\nu_\text{all}}^\text{CC}}} \ .
\end{equation}
As expected, $\sigma_{R_\text{exp}} \varpropto \left( T V_\text{eff} \right)^{-1/2}$, so that, for a given neutrino flux, the statistical error on $R_\text{exp}$ decreases with the time of exposure and the size of the detector.

\FIGURE[t!]{
 \scalebox{0.5}{\includegraphics{FLUXES.eps}} 
 \caption{High-energy astrophysical neutrino flux models as function of neutrino energy. The spectral index $\alpha=2.7$ for the Becker-Biermann (BB) and $2.3$, $2.6$ for the Koers-Tinyakov (KT) fluxes with and without source evolution, respectively. For the BB flux, we have set $\Gamma_\nu/\Gamma_\text{CR} = 3$ and $z_\text{CR}^{\max} = 0.03$ (see \cite{Becker:2008nf}). The atmospheric muon-neutrino flux, modelled according to \cite{Koers:2008hv}, which is considered as a background to the astrophysical neutrino signal, lies below the predictions of these models in the plotted energy range.}
 \label{FigFluxes}
}

\subsection{Astrophysical neutrino flux models}\label{Nu_prod}

\TABLE[h!]{
 \begin{tabular}{|l|c|c|c|}
  \hline
  Flux & $N_{\nu_\text{all}}^\text{NC}$ & $N_{\nu_\text{all}}^\text{CC}$ & $N_{\nu_\text{all}}^\text{NC}$/$N_{\nu_\text{all}}^\text{CC}$ \\
  \hline
  Waxman-Bahcall \cite{Waxman:2002wp}                                     & 1781.64  & 33.12   & 53.79 \\
  Becker-Biermann $\alpha=2.7$ \cite{Becker:2008nf}                       & 9248.58  & 130.68  & 70.77 \\
  Koers-Tinyakov no source evolution $\alpha=2.6$ \cite{Koers:2008hv}     & 9013.86  & 354.96  & 25.39 \\
  Koers-Tinyakov strong source evolution $\alpha=2.3$ \cite{Koers:2008hv} & 16495.92 & 1866.42 & 8.84 \\
  \hline
 \end{tabular}
 \caption{Expected number of neutral-current and charged-current events (summed over all flavours) at IceCube, $N_{\nu_\text{all}}^\text{NC}$ and $N_{\nu_\text{all}}^\text{CC}$, respectively, in the energy range $10^6 \le E_\nu/\text{GeV} \le 10^{12}$, for different choices of the incoming astrophysical neutrino flux. The exposure time used was $T = 15$ yr, and the effective detector volume, $V_\text{eff} = 1$ km$^3$. Only downgoing events are considered. The Waxman-Bahcall flux assumes an $E^{-2}$ spectrum, while for the Becker-Biermann and the Koers-Tinyakov fluxes we have used a power law of the form $E^{-\alpha}$, with $\alpha$ specified for each model. The different values of $\alpha$ have been selected so that the upper bound on the diffuse astrophysical muon-neutrino flux set by AMANDA-II \cite{Achterberg:2007qp} is satisfied. Details of the neutrino production models can be found in the indicated references.\label{TblNsh}}
}

We will present our estimations of $N_{\nu_\text{all}}^\text{NC}$, $N_{\nu_\text{all}}^\text{CC}$, and $N_{\nu_\text{all}}^\text{NC}/N_{\nu_\text{all}}^\text{CC}$, in the context of four models of astrophysical neutrino fluxes. The Waxman-Bahcall \cite{Waxman:2002wp} model makes use of the observation of ultra-high-energy ($> 10^{19}$ eV) cosmic rays to set an upper limit on the neutrino flux. The limit depends on the redshift evolution of the neutrino sources, which could be active galactic nuclei (AGN) or gamma-ray bursts. We have adopted, conservatively,
\begin{equation}
 \Phi_{\nu_\text{all}}^\text{WB}\left(E_\nu\right) = 10^{-8} \left(E_\nu/\text{GeV}\right)^{-2} ~\text{GeV}^{-1} ~\text{cm}^{-2} ~\text{s}^{-1} ~\text{sr}^{-1} \ .
\end{equation}
The second model, by Becker-Biermann \cite{Becker:2008nf}, describes the production of neutrinos in the relativistic jets of FR-I galaxies (low-luminosity radio galaxies with extended radio jets) through the decay of pions produced in the interaction of shock-accelerated protons with the surrounding photon field. The sources are assumed to evolve with redshift according to certain luminosity functions. The flux is given by
\begin{equation}
 \Phi_{\nu_\text{all}}^\text{BB}\left(E_\nu\right) \simeq 5.4 \times 10^{-3} \left(E_\nu/\text{GeV}\right)^{-2.7} ~\text{GeV}^{-1} ~\text{cm}^{-2} ~\text{s}^{-1} ~\text{sr}^{-1} \ .
\end{equation}
The third and fourth models, by Koers-Tinyakov \cite{Koers:2008hv}, predict an astrophysical neutrino flux based on the assumption that the neutrinos originate predominantly at AGN and that these behave like Centaurus A, the nearest active galaxy. The difference between these last two models lies in the assumption about the redshift evolution of the source: in one of them, sources are assumed not to evolve with redshift, while in the other one, they are assumed to have a strong redshift evolution, following the star-formation rate of $\sim\left(1+z\right)^3$. These two fluxes are given, respectively, by
\begin{eqnarray}
 \label{EqFluxKTevol}
 \Phi_{\nu_\text{all}}^\text{KT, no evol.}\left(E_\nu\right) 
 &\simeq& 3.5 \times 10^{-10} \left(E_\nu/\text{GeV}\right)^{-1.6} \times \nonumber \\
 && \times \min\left(1,E_\nu/E_{\nu,\text{br}}\right) ~\text{GeV}^{-1} ~\text{cm}^{-2} ~\text{s}^{-1} ~\text{sr}^{-1} \\
 \Phi_{\nu_\text{all}}^\text{KT, evol.}\left(E_\nu\right) 
 &\simeq& 4.6 \times 10^{-12} \left(E_\nu/\text{GeV}\right)^{-1.3} \times \nonumber \\
 && \times \min\left(1,E_\nu/E_{\nu,\text{br}}\right) ~\text{GeV}^{-1} ~\text{cm}^{-2} ~\text{s}^{-1} ~\text{sr}^{-1} \ ,
\end{eqnarray}
with $E_{\nu,\text{br}} = 4 \times 10^6$ GeV the break energy. Both the Becker-Biermann and Koers-Tinyakov models account for the change in cosmic-ray energy due to the adiabatic cosmological expansion, but only the latter take into account also energy losses due to pion photoproduction and electron-positron pair production in the interaction with the CMB photons. Currently, the most stringent upper bound on the astrophysical muon-neutrino flux is the one obtained by the AMANDA-II experiment \cite{Achterberg:2007qp}, which restricts the integrated muon-neutrino flux to be lower than $7.4 \times 10^{-8}$ GeV cm$^{-2}$ s$^{-1}$ sr$^{-1}$, within the interval 16 TeV \--- 2.5 PeV. We have checked that the four fluxes used in our analysis satisfy this upper bound in the standard-oscillation case, i.e., when the detected flavour ratios are $\phi_e:\phi_\mu:\phi_\tau = 1:1:1$. The spectral indices of the power laws for the Koers-Tinyakov fluxes, $-1.6$ and $-1.3$, have been chosen so that the integrated fluxes between 16 TeV and 2.5 PeV yield exactly the upper bound value set by AMANDA. A plot of the four different fluxes is presented in Fig.~\ref{FigFluxes}.

We have evaluated Eqs.~(\ref{EqNEvtsCC}) and (\ref{EqNEvtsNC2}) numerically in the range $10^6 \le E_\nu/\text{GeV} \le 10^{12}$ for the four different fluxes, assuming $T = 15$ yr and $V_\text{eff} = 1$ km$^3$ (the IceCube effective volume). The results are presented in Table \ref{TblNsh}. The Waxman-Bahcall model yields the lowest number of CC and NC events, while the Koers-Tinyakov model with strong source evolution yields the highest number, more than two orders of magnitude over Waxman-Bahcall. 

\subsection{Results}\label{result_exp}

\EPSFIGURE[t!]{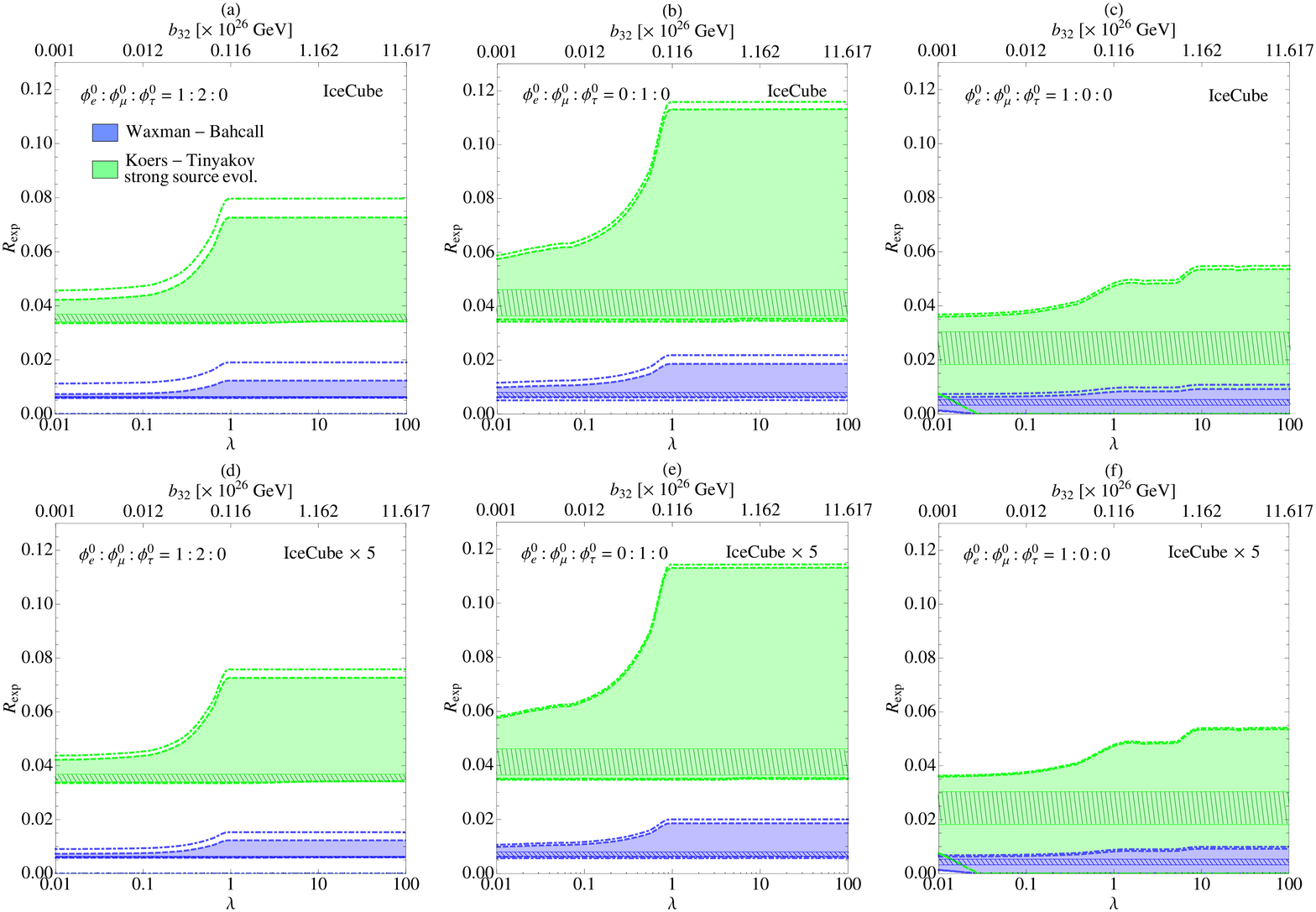,width=15.5cm}
 {$R_\text{exp} \equiv N_{\nu_\mu}/N_\text{sh}$ vs.~$\lambda$ for two neutrino flux models: the Waxman-Bahcall flux (in blue) and the Koers-Tinyakov flux with spectral index $\alpha = 2.3$ and strong source evolution (in green). Three initial flavour ratios have been considered: $\phi_e^0:\phi_\mu^0:\phi_\tau^0 = 1:2:0$ (plots (a) and (d)), $0:1:0$ (plots (b) and (e)) and $1:0:0$ (plots (c) and (f)). The coloured areas, bounded by dashed coloured lines, have been calculated plugging the values of $\tilde{\phi}_\mu$ from Fig.~\ref{Fig_OnlyPhimu} into Eq.~(\ref{EqRexp}) and using the event count data in Table \ref{TblNsh}. The dot-dashed coloured lines mark the $1\sigma$ uncertainty on $R_\text{exp}$ for each assumption of initial ratios, calculated with Eq.~(\ref{EqSigmaRexp}). For comparison, the standard value of $R_\text{exp}$ for the different initial ratios are shown in hatched bands of the corresponding colours; these values are calculated assuming no CPTV effects and using the current $3\sigma$ bounds on the standard mixing angles. In the upper row, an effective detector area $V_\text{eff} = 1$ km$^3$ (IceCube size) was assumed; for the bottom one, $V_\text{eff} = 5$ km$^3$ (IceCube $\times$ 5).\label{FigRexpvsLambda_WB_KTevol}}

In this section, we will display the regions of $R_\text{exp}$ vs.~$\lambda$, allowing for CPTV, for the Waxman-Bahcall model and the Koers-Tinyakov model with strong source evolution, which correspond, respectively, to the cases with the lowest and highest event yields. In order to generate these regions we have used the values in Table \ref{TblNsh} to calculate $R_\text{exp}$ in the presence of CPTV, Eq.~(\ref{EqRexp}), using for $\tilde{\phi}_\mu$ the range of values allowed by the variation of the standard mixing angles $\theta_{ij}$ within their current $3\sigma$ experimental bounds and of the new mixing angles $\theta_{bij}$ in the range $\left[0,\pi\right]$. This range of values of $\tilde{\phi}_\mu$ vs.~$\lambda$ can be obtained from the plot of $\phi_\mu$ vs.~$\lambda$ shown in Fig.~\ref{Fig_OnlyPhimu}, after applying the transformation in Eq.~(\ref{NormPhi}). The resulting regions as functions of $\lambda$ are shown colour-filled in Figure \ref{FigRexpvsLambda_WB_KTevol}, for the Waxman-Bahcall (blue) and Koers-Tinyakov (green) models. We have explored two different detector effective volumes, $1$ km$^3$ (IceCube-sized) and 5 km$^3$, and the three different choices for the initial fluxes $\phi_e^0:\phi_\mu^0:\phi_\tau^0 = 1:2:0$, $0:1:0$ and $1:0:0$, that we introduced in the previous sections. Furthermore, this figure includes boundaries of $1\sigma$ statistical uncertainty on $R_\text{exp}$ that were obtained by adding (subtracting) $1\sigma$ to (from) the upper (lower) boundaries of the coloured regions. The values of $\sigma$ were calculated by plugging into Eq.~(\ref{EqSigmaRexp}) the corresponding values of $\tilde{\phi}_\mu$ that occur on the borderlines of the coloured regions. 

We note that the shapes of the coloured regions in Fig.~\ref{FigRexpvsLambda_WB_KTevol} are similar to those in Fig.~\ref{Fig_OnlyPhimu}. This can be understood if we note that $R_\text{exp}$, Eq.~(\ref{EqRexp}), is proportional to $\tilde{\phi}_\mu$. In Fig.~\ref{Fig_OnlyPhimu} we saw that there exists overlap among the regions related to the different hypotheses of the initial flux. In fact, in Fig.~\ref{FigRexpvsLambda_WB_KTevol}, there is overlap between the three production models, with the region corresponding to $\phi_e^0:\phi_\mu^0:\phi_\tau^0 = 1:2:0$ almost entirely enclosed within the region for $0:1:0$. This last fact can be explained on account of the reduction in size of the region for $1:2:0$, caused by the normalisation of $\tilde{\phi}_\mu$ applied for this case, i.e., $\tilde{\phi}_\mu = \phi_\mu/3$. As we can anticipate, due to the higher event yield, the statistical uncertainty on $R_\text{exp}$ associated to the Koers-Tinyakov flux is lower than the one associated to the Waxman-Bahcall flux, and is reduced when the larger detector volume is used. The size of this uncertainty is also proportional to the value of $\tilde{\phi}_\mu$ (see Eq.~(\ref{EqSigmaRexp})). As a consequence, the size of the $1\sigma$ regions is larger for an initial flux of $1:2:0$, intermediate for $0:1:0$, and smallest for $1:0:0$. 

When a detector volume of 1 km$^3$ is considered, there is a clear overlap among regions corresponding to different assumptions of the neutrino flux model when the production scenario is $1:0:0$. This observation is reinforced when we consider also the regions spanned by the statistical uncertainty. In comparison, for the $1:2:0$ and $0:1:0$ scenarios, the regions corresponding to the two flux models do not overlap. When the 5 km$^3$ detector is assumed, the regions associated to $1:2:0$ and $0:1:0$ are further separated at the $1\sigma$ level. However, there is still an overlap between the two flux models in the $1:0:0$ scenario. Here we do not show the results for the Becker-Biermann model and the Koers-Tinyakov model with no source evolution, since their results are embodied in what we have already presented. For instance, the mean value of $R_\text{exp}$ and $\sigma_{R_\text{exp}}$ for Becker-Biermann is similar to the one for Waxman-Bahcall. Similar estimates can be easily obtained by plugging the values of Table \ref{TblNsh} into Eqs.~(\ref{EqRexp}) and (\ref{EqSigmaRexp}). 
              
\section{Summary and conclusions}\label{Section_Conclusions}

Motivated by the CPT-violating (CPTV) neutrino coupling considered in the Standard Model Extension, we have added a CPTV, energy-independent, contribution to the neutrino oscillation Hamiltonian and explored its effects on the flavour ratios of the high-energy (1 PeV and higher) astrophysical neutrino flux predicted to come from active galactic nuclei. We have parametrised the strength of the CPTV contribution by the parameter $\lambda$, defined as the quotient between the eigenvalues of the CPTV Hamiltonian, $b_{21}$ and $b_{32}$, and those of the standard-oscillation one, $\Delta m_{21}^2/\left(2E^\star\right)$ and $\Delta m_{32}^2/\left(2E^\star\right)$, with $E^\star = 1$ PeV, and allowed $\lambda$ to vary between $10^{-2}$ and $100$, corresponding to standard-oscillation dominance and CPTV dominance, respectively.  We have used three different neutrino production scenarios for the flavour ratios at the astrophysical sources: production by pion decay, which results in $\phi_e^0:\phi_\mu^0:\phi_\tau^0 = 1:2:0$; muon cooling, which results in $0:1:0$; and neutron decay, resulting in $1:0:0$, and explored the effect of a potential CPTV on the neutrino flavour ratios at Earth, $\phi_\alpha$ ($\alpha = e, \mu, \tau$), and on the ratios between them. With this objetive, we have studied the behaviour of $\phi_\mu$, $R = \phi_\mu/\phi_e$, and $S = \phi_\tau/\phi_\mu$, by letting the standard-oscillation mixing parameters vary within their current $3\sigma$ experimental bounds, and varying the unknown CPTV parameters as broadly as possible, while keeping $b_{21}$ and $b_{32}$ below their current upper limits (obtained from atmospheric and solar experiment data).

From the observation of $\phi_\mu$, if CPTV is dominant, we found that there could be large deviations with respect to the pure standard-oscillation case, depending on the values of the CPTV parameters. These deviations start at $\lambda = 0.1$ ($b_{32} \sim 10^{-28}$ GeV, $b_{21} \sim 10^{-26}$ GeV) and reach a plateau at $\lambda = 1$. There are overlaps between the different neutrino production models, in such a way that a measurement of certain values of $\phi_\mu$ could be satisfied by two different production models, either including CPTV or not.

When we consider the possibility of detecting $\phi_\mu$ and $\phi_e$, from which $R \equiv \phi_\mu/\phi_e$ can be built, we find that the regions corresponding to $1:2:0$ and $0:1:0$ exhibit a similar behaviour than when $\phi_\mu$ alone is measured, though the former is now nearly contained by the latter. This is not the case for $1:0:0$, where the value of $R$ could blow up owing to a potentially very low value of $\phi_e$, in comparison to $\phi_\mu$.

If tau-neutrinos can also be detected, then we can use the ratio $S \equiv \phi_\tau/\phi_\mu$.  When we combine $R$ and $S$, i.e., when we assume the ability to measure $\phi_\tau$ with enough statistics, we improve the chances of discovering CPTV effects. In fact, large CPTV regions of $\left(R,S\right)$ values that are well distinguished from the standard-oscillation case are obtained for the three neutrino production models that we have considered here. As a consequence of the wideness of these regions, there are many overlapping areas where a given pair $\left(R,S\right)$ can be generated by any of the production models that we have explored, assuming a dominant CPTV.

On the other hand, in the case of the non-observation of deviations in the flavour ratios, it will be possible to impose very stringent limits on the parameters related to CPTV in the neutrino sector, such as $b_{21} \lesssim 10^{-29}$ GeV and $b_{32} \lesssim 10^{-27}$ GeV, to be compared with the current limits of $10^{-23}$ GeV and $10^{-21}$ GeV, respectively. 

In order to compare CPTV with other competitive new physics scenarios, we have included the possibility of neutrino decay to invisible products, both in the normal and inverted mass hierarchies. As a result, there are additional overlaps between the CPTV regions and the regions accessible by neutrino decays, for certain values of the CPTV parameters.

With the purpose of presenting a more realistic perspective, we have performed an analysis of the potential CPTV signals at a large ice \u{C}erenkov detector such as IceCube. On top of the three neutrino production models, $1:2:0$, $0:1:0$ and $1:0:0$, we have used two different models for the neutrino  astrophysical flux, one by Waxman and Bahcall and the other by Koers and Tinyakov, and compared their respective signals at the detector. For this analysis, we have found that there are still overlaps among the three production models, even more pronounced than the ones that were obtained in the theoretical plots. In 15 years of exposure time, a 1 km$^3$ detector (or, equivalently, $t V_\text{eff} = 15$ yr km$^3$) would be able to distinguish between the two fluxes, in the $1:2:0$ and $0:1:0$ scenarios, while a 5 km$^3$ detector with the same exposure (or $t V_\text{eff} = 75$ yr km$^3$), would provide clearer separation between the flux models at the $1\sigma$ level.

When $\lambda \geq 1$, a separation of a few standard deviations between the CPT-conserving and the CPTV scenarios is possible, depending on the values of the CPTV mixing parameters, both for the Waxman-Bahcall and the Koers-Tinyakov fluxes. This separation is more clearly visible in the $1:2:0$ and $0:1:0$ cases. Our main result is that, while it is in principle possible to detect the presence of CPTV with IceCube, it will not be possible, for many values of $\left(R,S\right)$, to find which one of the production models is the actual one. The detection of CPTV can be aided by a more precise knowledge of the standard-oscillation mixing parameters and by a considerably higher event yield, brought by a larger effective volume or probably from a non-\u{Cerenkov} detector, but these two improvements do not eliminate the overlaps that exist between $\left(R,S\right)$ regions associated to different production models. In the event of measuring values of $R$, $S$, or both, that fall inside an overlap region, then, in the absence of knowledge of the values of the CPTV mixing parameters, one is able to trade one production model, with a certain set of values of the CPTV parameters, by one of the other overlapping models, with another set of values of the parameters. These degeneracies could be lifted by an independent measurement of the CPTV parameters in the neutrino sector, in a different kind of experiment.

Nevertheless, as a tool for detecting the presence of CPT breaking in the neutrino sector, if not for measuring it in detail, IceCube might be a useful one. If CPT is broken, and if the parameters introduced by the breaking have certain values, then IceCube could be able to detect the deviation from the standard-oscillation scenario after 15 years of data taking.

\acknowledgments

This work was supported by grants from the Direcci\'on Acad\'emica de Investigaci\'on of the Pontificia Universidad Cat\'olica del Per\'u (projects DAI-4075 and DAI-L009 [LUCET]) and by a High Energy Latinamerican-European Network (HELEN) STT grant. MB acknowledges the hospitality of IFIC during the development of this work.

\end{document}